\documentclass[a4paper]{aa}

\usepackage{graphicx,float}
\usepackage{latexsym,amsmath,amssymb}
\usepackage{natbib}
\bibpunct{(}{)}{;}{a}{}{,}

\def \arvind    {A.~N. Parmar}

\def \laurence  {L. Boirin}

\def \mariano {M. M{\'e}ndez}
\def \maria {M. D{\'i}az Trigo}
\def \jelle {J.~S. Kaastra}

\def \sron {SRON, National Institute for Space Research,
  Sorbonnelaan 2, 3584 CA  Utrecht, The Netherlands}

\def \estec {Research and Scientific Support
                Department of ESA, ESTEC, Postbus 299, 2200 AG
                Noordwijk, The Netherlands}

\def \stras {Observatoire Astronomique de Strasbourg, 11 rue de
                 l'Universit\'e, F-67000 Strasbourg, France}

\newcommand {\catwenty} {\ion{Ca}{xx}}
\newcommand {\fenineteen} {\ion{Fe}{xix}}

\newcommand {\oseven} {\ion{O}{vii}}
\newcommand {\oeight} {\ion{O}{viii}}

\newcommand {\netwo} {\ion{Ne}{ii}}
\newcommand {\nenine} {\ion{Ne}{ix}}
\newcommand {\neten} {\ion{Ne}{x}}

\newcommand {\niteight} {\ion{Ni}{xxviii}}

\def\ang {$\rm\AA$}

\def\degmark{^\circ}
\def \rsun {\ifmmode$R$_{\odot}\else R$_{\odot}$}
\newcommand{\mc}{\multicolumn}

\def \flux {$F_{\rm X}$}

\def \hcm {\hbox {\ifmmode $ atoms cm$^{-2}\else atoms cm$^{-2}$\fi}}

\def\approxgt{\mathrel{\hbox{\rlap{\lower.55ex \hbox {$\sim$}}
        \kern-.3em \raise.4ex \hbox{$>$}}}}
\def\approxlt{\mathrel{\hbox{\rlap{\lower.55ex \hbox {$\sim$}}
        \kern-.3em \raise.4ex \hbox{$<$}}}}

\def\arcsec{\hbox{$^{\prime\prime}$}}

\newcommand {\ergcms} {erg cm$^{-2}$ s$^{-1}$}
\newcommand {\chisq} {$\chi ^{2}$}
\newcommand {\rchisq} {$\chi_{\nu} ^{2}$}

\newcommand {\phind} {$\Gamma$}

\newcommand {\egau} {$E_{\rm gau}$}

\newcommand {\ktbb} {$kT_{\rm bb}$}

\newcommand {\ew} {$EW$}
\newcommand {\ews} {$EW$s}

\def \nh {$N{\rm _H}$}
\def \nhabs {$N{\rm _H^{abs}}$}
\def \nhxabs {$N{\rm _H^{xabs}}$}

\newcommand {\ttnh} {$\times~$10$^{22}$~atom~cm$^{-2}$}

\newcommand {\fetfive} {\ion{Fe}{xxv}}
\newcommand {\fetsix} {\ion{Fe}{xxvi}}

\newcommand {\lya} {Ly$\alpha$}
\newcommand {\lyb} {Ly$\beta$}

\def \xiunit {\hbox{erg cm s$^{-1}$}}
\def \logxi {$\log(\xi)$}
\def \xil {$\xi$}

\def \xabs {{\tt xabs}}
\newcommand {\sigmav} {$\sigma_{\rm v}$}
\newcommand {\kms} {km~s$^{-1}$}

\def \kpl {$k_{\rm pl}$}
\def \kbb {$k_{\rm bb}$}
\def \kgau {$k_{\rm gau}$}

\def \nineteen {XB\,1916$-$053}
\def \mxb {MXB\,1659$-$298}
\def \bigdip {X\,1624$-$490}
\def \twelve {XB\,1254$-$690}
\def \grs {GRS\,1915+105}
\def \gro {GRO\,J1655$-$40}
\def \gx {GX\,13+1}

\def \exo {EXO\,0748$-$676}

\def \thirteen {XB\,1323$-$619}
\def \seventeen {H\,1743$-$322}
\def \threethreenine {GX\,339$-$4}
\def \xte {XTE\,J1650$-$500}
\def \fouru {4U\thinspace1746$-$371}

\begin{document}

\title{Spectral changes during dipping in low-mass X-ray
binaries due to highly-ionized absorbers}

\author{\maria\inst{1} \and \arvind\inst{1} \and \laurence\inst{2}
\and \mariano\inst{3} \and \jelle\inst{3}}

\offprints{M. D{\'i}az Trigo \email{mdiaz@rssd.esa.int}}

\institute{\estec \and \stras \and \sron}

\date{Received 7 June 2005 / Accepted 29 August 2005}

\authorrunning{M. D{\'i}az Trigo et al.}

\titlerunning{Highly-ionized absorbers in LMXBs}

\abstract{X-ray observations have revealed that many microquasars and
 low-mass X-ray binaries (LMXBs) exhibit narrow absorption features
 identified with resonant absorption from \fetfive\ and \fetsix\ and
 other abundant ions. In many well studied systems there is evidence
 for blue-shifts, indicating outflowing plasmas. We succesfully model
 the changes in \emph{both} the X-ray continuum and the Fe absorption
 features during dips from all the bright dipping LMXBs observed by
 XMM-Newton (\exo, \twelve, \bigdip, \mxb, \fouru\ and \nineteen) as
 resulting primarily from an increase in column density and a decrease
 in the ionization state of a highly-ionized absorber in a similar way
 as was done for \thirteen. This implies that the complex spectral
 changes in the X-ray continua observed from the dip sources as a
 class can be most simply explained primarily by changes in the highly
 ionized absorbers present in these systems. There is no need to
 invoke unusual abundances or partial covering of extended emission
 regions. Outside of the dips, the absorption line properties do not
 vary strongly with orbital phase. This implies that the ionized
 plasma has a cylindrical geometry with a maximum column density close
 to the plane of the accretion disk.  Since dipping sources are simply
 normal LMXBs viewed from close to the orbital plane this implies that
 ionized plasmas are a common feature of LMXBs.  } \maketitle

\section{Introduction}
\label{sec:intro}

Around 10 galactic low-mass X-ray binaries (LMXBs) exhibit
periodic dips in their X-ray intensity. The dips recur at the
orbital period of the system and are believed to be caused by
periodic obscuration of a central X-ray source by structure
located in the outer regions of a disk resulting from the impact
of the accretion flow from the companion star into the disk
\citep{1916:white82apjl}. The depth, duration and spectral
properties of the dips vary from source to source and from cycle
to cycle. The spectral changes during LMXBs dips are complex and
cannot be well described by a simple increase in column density of
cold absorbing material with normal abundances
\citep[e.g.,][]{1254:courvoisier86apj, 1916:smale92apj}.

Modeling of these spectral changes provides a powerful means of
studying the structure and location of the emitting and absorbing
regions in LMXBs.
 Two approaches have been used. Initially, in the ``absorbed plus unabsorbed''
approach \citep[e.g., ][]{0748:parmar86apj} the persistent
(non-dipping) spectral shape was used to model spectra from
dipping intervals. The spectral evolution during dipping was accounted
for by a large increase in the column density of the absorbed
component, and a decrease of the normalization of the unabsorbed
component. The latter decrease was attributed to electron
scattering in the absorber.  More recently, in the ``complex
continuum'' approach \citep[e.g.,][]{1624:church95aa}, the X-ray
emission is assumed to originate from a point-like blackbody, or
disk-blackbody component, together with an extended power-law
component. This approach primarily models the spectral changes
during dipping intervals by the partial and progressive covering
of the power-law emission from an extended source. The
absorption of the point-like component is allowed to vary
independently from that of the extended component, and usually no partial
covering is included.

The improved sensitivity and spectral resolution of {\it Chandra}
and XMM-Newton is allowing narrow absorption features from highly
ionized Fe and other metals to be observed from a growing number
of X-ray binaries.  These features were first detected from the
micro-quasars \gro\ \citep{1655:ueda98apj,1655:yamaoka01pasj} and
\grs\ \citep{1915:kotani00apj,1915:lee02apj}.  More recent {\it
Chandra} High-Energy Transmission Grating Spectrometer (HETGS)
observations of the black hole candidate \seventeen\ \citep{1743:miller04apj}
revealed the presence of blue-shifted \fetfive\ and \fetsix\
absorption features indicative of a highly-ionized outflow. The
LMXB systems that exhibit narrow X-ray absorption features are all
known dipping sources \citep[see Table~5 of][]{1916:boirin04aa}
except for \gx.
 This source shows deep blue-shifted Fe
absorption features in its HETGS spectrum, again indicative of
outflowing material \citep{gx13:ueda04apj}.

\citet{1323:boirin05aa} examined the changes in the equivalent widths,
$EW$s, of the \fetfive\ and \fetsix\ absorption features seen from
\thirteen\ during persistent and dipping intervals. They found
evidence for the presence of less-ionized material in the line of
sight during dips consistent with a decrease in the photo-ionization
parameter, \logxi, from $3.9 \pm 0.1$ \xiunit\ to $3.13 \pm 0.07$
\xiunit, while the equivalent hydrogen column density, \nhxabs, of the
ionized absorber increased from ($3.8 \pm 0.4$)~\ttnh\ to ($37 \pm
2$)~\ttnh. There was also a smaller increase in the equivalent
hydrogen column of neutral material, \nh, from $(3.50 \pm 0.02) \times
10^{22}$~atom~cm$^{-2}$ to $(4.2 \pm 0.2) \times
10^{22}$~atom~cm$^{-2}$. During persistent intervals almost all the
abundant elements, except for Fe, are fully ionized, whilst during
dips this is not the case. This results in the presence of absorption
edges and many additional absorption features that blend together
causing an apparent change in the 1--10~keV continuum measured by
XMM-Newton.  \citet{1323:boirin05aa} went on to demonstrate that the
changes in the source 1--10~keV \emph{continuum} during dips are
consistent with being due primarily to changes in the ionized absorber
and that there is no need to invoke unusual abundances or complex
continuum models to explain them.

Similar changes in the \fetsix\ and \fetfive\ absorption line $EW$s
during dips have been reported from \bigdip\
\citep{1624:parmar02aa} and \nineteen\ \citep{1916:boirin04aa}. In
the case of \mxb, whilst \citet{1658:sidoli01aa} report that there
is no obvious orbital dependence of the $EW$s of the Fe absorption
features, their Fig.~4 shows that the ratio of \fetfive/\fetsix\
$EW$s is at a maximum during dips, consistent with the presence of
less-ionized material. As pointed out
by \citet{1323:boirin05aa}, since highly-ionized absorption
features are seen from many other dip sources, changes in ionized
absorbers may also explain the overall variations in X-ray spectra
observed during LMXB dipping intervals as a class. Here, we
examine this prediction by generalizing the analysis performed on
\thirteen\ to all the other bright dipping sources observed by
XMM-Newton: \exo, \twelve, \bigdip, \mxb, \fouru, and \nineteen.
The overall properties of these sources are given in
Table~\ref{tab:lmxb-prop}, while the previously reported
XMM-Newton results are summarized in Table~\ref{tab:lines}.
\thirteen\ is included for completeness and comparison. The
previously bright dip source 4U\,1755$-$338 was observed by
XMM-Newton \citep{1755:angelini03apj}, but is too faint for a
detailed spectral study and is not discussed further.

\begin{table}
\caption{Overall properties of the LMXBs studied here. $L_{36}$ is the
best-fit 0.6--10~keV unabsorbed luminosity in units of
10$^{36}$~erg~s$^{-1}$ for distances $d$, except for
\thirteen\ where the 0.5--10~keV unabsorbed luminosity is from
\citet{1323:boirin05aa}. The orbital periods, $P_{\rm orb}$, are from
\citet[][]{RitterKolb03aa}. D indicates the presence of dips and
E, eclipses.} \label{tab:lmxb-prop}
\begin{center}
\begin{tabular}[c]{lcccc}
\hline \hline\noalign{\smallskip}
LMXB &  $P_{\rm orb}$ & $L_{36}$ & $d$ & Dips/ \\
& (hr) & (erg s$^{-1}$) & (kpc) & Eclipses \\
\noalign{\smallskip\hrule\smallskip}
\nineteen & 0.8 & 4.4 & 9.3 & D \\
\thirteen & 2.9 & 5.2 & 10 & D  \\
\exo & 3.8 & 3.4  & 10 & D,E  \\
\twelve & 3.9 & 10.4 & 10 & D   \\
\fouru & 5.8 & 10.1 & 10.7 & D   \\
\mxb & 7.1 & 34.4 & 15 & D,E  \\
\bigdip & 21 & 47.5 & 15 & D  \\
\noalign{\smallskip\hrule\smallskip}
\label{tab:lmxb-prop}
\end{tabular}
\end{center}

\end{table}

\begin{table*}
\caption{Previously reported emission and absorption features near
7~keV from the LMXBs studied here. $E{\rm _{gau}}$ and $EW$
are the energy and equivalent width of the best-fit Gaussian
profiles. The continuum model on which the features are
superposed is indicated. {\tt dbb}, {\tt bb}, {\tt pl}, and {\tt
cpl} indicate disk-blackbody, blackbody, power-law and cutoff
power-law models, respectively.}
\begin{center}
\begin{tabular}{l@{\extracolsep{0.cm}}l@{\extracolsep{0.15cm}}l@{\extracolsep{0.15cm}}l@{\extracolsep{0.15cm}}l@{\extracolsep{0.15cm}}l@{\extracolsep{0.1cm}}l@{\extracolsep{0.1cm}}l@{\extracolsep{0.1cm}}}
\hline \hline\noalign{\smallskip}
LMXB & Continuum & \multicolumn{5}{c}{Gaussian features near 7 keV} & Reference \\
& model & \multicolumn{3}{c}{Absorption} & \multicolumn{2}{c}{Emission} & \\
\noalign{\smallskip\hrule\smallskip}
 & & Ident. & \egau\ (keV) & \ew\ (eV) & \egau\ (keV) & \ew\ (eV) & \\
\noalign{\smallskip}
\nineteen\ & dbb+pl & \fetfive & 6.65$^{+0.05}_{-0.02}$ & 30$^{+12}_{-8}$ &  &  & \citet[][]{1916:boirin04aa} \\
 & & \fetsix & 6.95$^{+0.05}_{-0.04}$ & 30$^{+12}_{-11}$ &  &  & \\
\noalign{\smallskip}
\thirteen\ & bb+pl & \fetfive & 6.68 $\pm$ 0.04 & 25$^{+19}_{-7}$  & 6.6$^{+0.1}_{-0.2}$ & & \citet[][]{1323:boirin05aa} \\
& & \fetsix & 6.97 $\pm$ 0.05 & 24$^{+21}_{-7}$ & &  &  \\
\noalign{\smallskip}
\twelve\ & dbb+pl & \fetsix~Ly$\alpha$ & 6.95 $\pm$ 0.03 & 27$^{+8}_{-11}$ &  & & \citet[][]{1254:boirin03aa} \\
& & \fetsix~Ly$\beta$ & 8.20$^{+0.05}_{-0.10}$  &17 $\pm$ 9 &  &  & \\
\noalign{\smallskip}
\mxb\ \,   & bb+cpl & \fetfive & 6.64 $\pm$ 0.02 & 33$^{+20}_{-9}$ & 6.47$^{+0.18}_{-0.14}$ & 160$^{+60}_{-40}$ & \citet[][]{1658:sidoli01aa} \\
    &   & \fetsix & 6.90$^{+0.02}_{-0.01}$ & 42$^{+13}_{-8}$ &  &  & \\
\noalign{\smallskip}
\bigdip\ & bb+pl & \fetfive & 6.72 $\pm$ 0.03 & 7.5$^{+6.3}_{-1.7}$ & 6.58$^{+0.07}_{-0.04}$ & 78$^{+19}_{-6}$  & \citet[][]{1624:parmar02aa} \\
& & \fetsix & 7.00 $\pm$ 0.02 & 16.6$^{+5.9}_{-1.9}$ &  &  & \\
\noalign{\smallskip\hrule\smallskip}
\label{tab:lines}
\end{tabular}
\end{center}
\label{tab:lmxb-lines}
\end{table*}

\section{Data analysis}
\label{sec:reduction}

The XMM-Newton Observatory \citep{jansen01aa} includes three
1500~cm$^2$ X-ray telescopes each with an European Photon Imaging
Camera (EPIC) at the focus. Two of the EPIC imaging spectrometers use
MOS CCDs \citep{turner01aa} and one uses pn CCDs
\citep{struder01aa}. Reflection Grating Spectrometers
\citep[RGS,][]{denherder01aa} are located behind two of the
telescopes. Since the pn has an effective area a factor of $\sim$5
higher at 7~keV and a better energy resolution than the MOS CCDs, we
concentrate on the analysis of pn data. Table~\ref{tab:obslog} gives
details of the observations and modes used. The pn was operated in
either Timing or Small Window mode. We obtained all data products from
the XMM-Newton public archive and reduced them using the Science
Analysis Software (SAS) version 6.0.0 and calibration files released
on 30 June 2004, except for (1) the \twelve\ data, which were reduced
using SAS version 5.4.1 and (2) the \exo\ data which were reduced with
version 6.1.0. In order to minimize the effect of any background
variations on the extracted spectra, we excluded from the analysis
intervals where the $>$10~keV pn count rate (one CCD) was $>$1
s$^{-1}$, except for \twelve\ (see Sect. \ref{sec:twelve}).

\begin{table*}[ht!]
\caption{Observation log for the EPIC pn camera. $T$ is the net
exposure time and $CR$ is the pn 0.6--10 keV persistent emission
count rate. SW indicates Small Window mode.}
\begin{center}
\begin{tabular}[c]{lccllll}
\hline \hline\noalign{\smallskip}
 LMXB
&  Start time   &  \mc{1}{c}{$T$}
& Mode & Filter & $CR$ \\
     & (yr~mon~dy~hr:mn)   &
\mc{1}{c}{(ks)}  & & & (s$^{-1}$)  \\
\noalign{\smallskip\hrule\smallskip}
\nineteen\
           & 2001 Sep 25 04:14  & 16.5 &  Timing     & Medium & 75  \\
\exo\
           & 2003 Nov 12 08:23  & 94.6 & SW     & Medium & 30  \\
\twelve\
           &  2001 Jan 22 15:49  & 17.6 & Timing     & Thin1 & 190  \\
\fouru\
& 2002 Sep 19 11:59  & 45.6 & SW & Medium & 110 \\
\mxb\
           & 2001 Feb 20 08:28 &  31.6 &  SW     & Thin1 & 160  \\
\bigdip\
           & 2001 Feb 12 07:29   & 57.5 &  SW  & Medium & 75   \\
\noalign{\smallskip\hrule\smallskip} \label{tab:obslog}
\end{tabular}
\end{center}
\label{tab:obslog}
\end{table*}

In pn Timing mode, only one CCD chip (corresponding to a field of view
of 13\farcm6$\times$4\farcm4) is used and the data from that chip are
collapsed into a one-dimensional row (4\farcm4) to be read out at high
speed, the second dimension being replaced by timing information. This
allows a time resolution of 30~$\mu$s, and photon pile-up occurs only
for count rates $>$1500 s$^{-1}$, much brighter than any of the
sources analyzed here. Pile-up occurs when two or more photons hit the
same, or adjacent, pixels during a single CCD read-out cycle, and is
accounted for as a single event with an energy that is the sum of the
energies of the individual photons. Pile-up reduces the observed flux
of a source and makes the spectrum artificially harder than it
actually is.  Only single and double events (patterns 0 to 4) were
selected to reduce pile-up effects.  Source events were extracted from
53\arcsec\ wide columns centered on the source positions. Background
events were obtained from columns of the same width, but centered well
away from the sources.

In pn Small Window mode, only a 63$\times$63 pixel region of the
central CCD is read out every 5.7~ms.  The count rates of the sources
observed in Small Window mode are close to, or above, the 100 s$^{-1}$
level, above which pile-up effects become important
\citep{struder01aa}. Multiple events that are incorrectly assigned as
single should not produce additional line features. We used the SAS
task {\tt epatplot}, which utilizes the relative ratios of single- and
double-pixel events which deviate from standard values in case of
significant pile-up, as a diagnostic tool in the pn camera Small
Window mode data. As expected, this showed that the persistent spectra
extracted for all the sources observed in Small Window mode were
affected by pile-up.  We investigated further the importance of
properly correcting for count pile-up in Small Window mode by
extracting events in annuli of inner radii 15, 20, and 25\arcsec\ and
an outer radius of 30\arcsec\ centered on the PSF core for \mxb. This
source was chosen because it has the highest count rate in Small
Window mode (see Table~\ref{tab:obslog}).  We compared then the
results of spectral fits to those obtained when the PSF core was
included. We obtained consistent spectral results once events within a
radius of 15\arcsec\ were excluded. We repeated a similar process for
all the sources affected by pile-up and excluded events from the inner
9\farcs25, 17\farcs5, and 15\arcsec\ radius core of the point spread
function (PSF) from the pn spectra of \exo, \fouru\ and \mxb,
respectively.  In order not to produce any additional uncertainties,
we used the same extraction regions for the dip spectra for these
sources.

\begin{figure*}[ht!]
\centerline{\includegraphics[angle=0,width=1.05\textwidth]{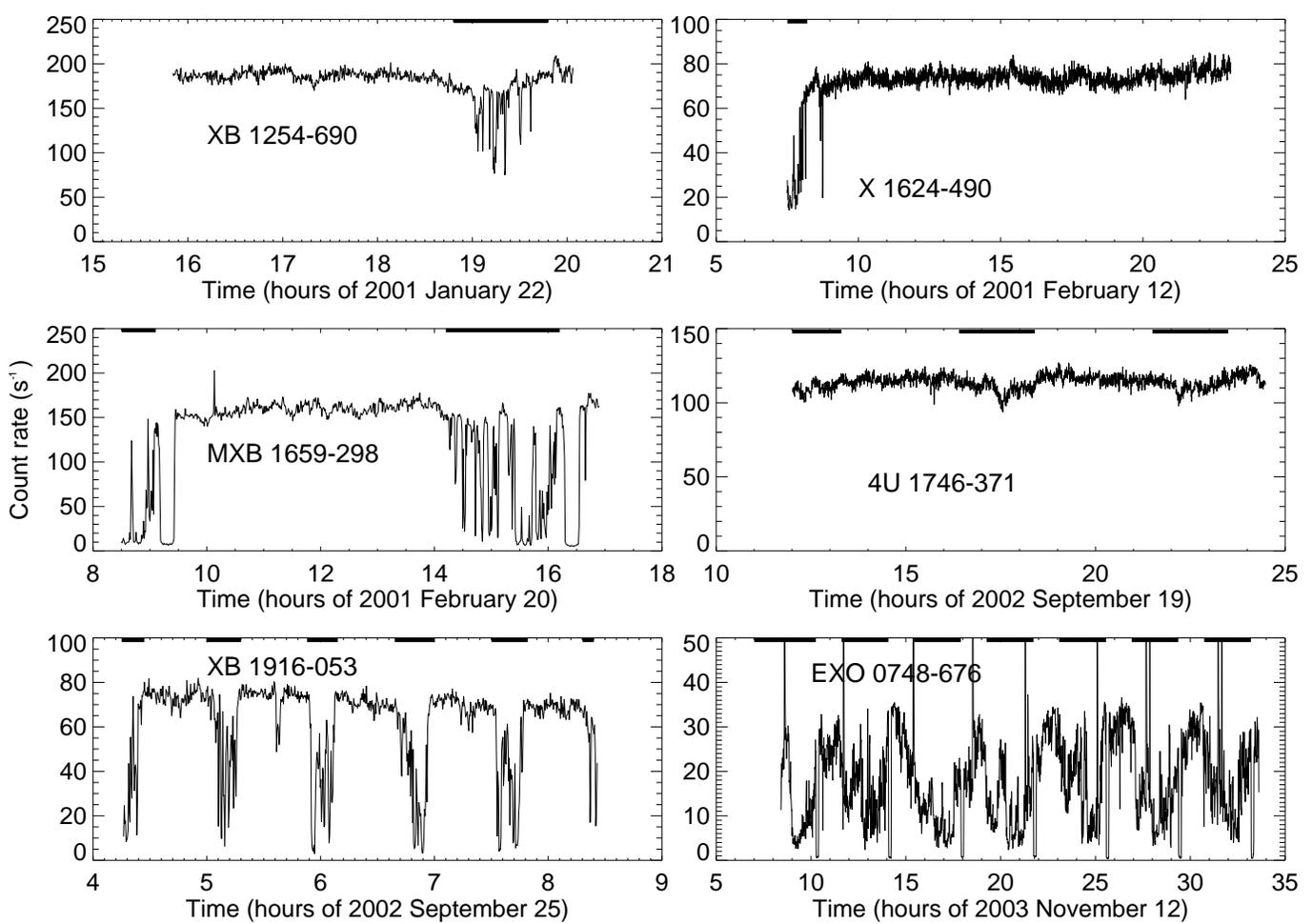}}
\caption{EPIC pn 0.6--10 keV lightcurves. The
binning is 20~s except for \mxb\ and \exo, where it is
40~s. The thick horizontal lines mark the intervals used to
extract dip spectra.} \label{fig:lightcurves}
\end{figure*}

Figure~\ref{fig:lightcurves} shows the EPIC pn 0.6--10~keV lightcurves
of the sources studied here. All the sources except for \fouru\ show
deep dipping activity. Regular eclipses are visible from \exo\ and
\mxb. XMM-Newton has made a number of observations of \exo\ and we
chose the 2002 November 12 observation since the dips appeared to be
unusually prominent.  \bigdip\ appears to be already in a deep dip at
the start of the observation, and only one dip is seen from \twelve\
close to the end of the observation.  We first excluded X-ray bursts
and eclipses from the lightcurves and then selected intervals flagged
with thick horizontal lines in Fig.~\ref{fig:lightcurves} for the
extraction of dip spectra and dip-free intervals for the persistent
emission spectra.  Within the dipping intervals, we extracted between
1 and 5 spectra for each source based on the intensity selection
criteria given in Table~\ref{tab:dipselection}. \exo\ and \mxb\
spectra with dip depths $>$80\% of the persistent emission were
excluded from the analysis since they are strongly influenced by
background counts.

\begin{table}[ht!]
\begin{center}
\caption{Dip intensity selection. Dip depth is given as a percentage
of the 0.6--10~keV pn persistent count rate (2--10~keV for \bigdip). The
accumulation time was 20~s.} \begin{tabular}{cc}
\hline\hline\noalign{\smallskip}
Dip Level & Dip depth\\
 \noalign{\smallskip\hrule\smallskip}
1 & $<$15 \% \\
2 & 15--30 \%  \\
3 & 30--50 \%  \\
4 & 30--65 \% \\
5 & 65--80 \%  \\
\noalign{\smallskip\hrule\smallskip}
\label{tab:dipselection}
\end{tabular}
\end{center}
\end{table}

\section{Spectral results}
\label{sec:spectralanalysis}

\begin{table*}[ht!]
\caption{Ionizing continua used to pre-calculate grids of relative
ionic column densities for fitting with the \xabs\ model in SPEX (see
text). The references are for the cutoff energies, $E_c$.}  \begin{center}
\begin{tabular}{lcccl}
\hline \hline\noalign{\smallskip}
LMXB & $\Gamma$ & $E_{c}$ (keV) & Mission & Reference\\
\hline\noalign{\smallskip}
\nineteen &1.87& 80 & BeppoSAX & \citet[][]{1916:church98aa} \\
\thirteen &1.96& 44 & BeppoSAX & \citet[]{1323:balucinska99aa}\\
\exo &1.27 & 44 & BeppoSAX & \citet{0748:sidoli05aa}\\
\twelve &1.28 & 5.9 & R--XTE & \citet[][]{1254:smale02apj}\\
\fouru & 0.39 & 3.5 & BeppoSAX & \citet[][]{1746:parmar99aa} \\
\mxb &1.21 & 7.1& BeppoSAX & \citet{1658:oosterbroek01aa}\\
\bigdip &2.00& 12 & BeppoSAX & \citet{1624:balucinska00aa} \\
\noalign{\smallskip\hrule\smallskip}
\label{tab:ioniz-cont}
\end{tabular}
\end{center}
\label{tab:ion-con}
\end{table*}

We performed spectral analysis using XSPEC \citep{arnaud96conf}
version 11.2, and SPEX \citep{kaastra96} version 2.00.11. We used
throughout the photo-electric cross sections of \citet{morrison83apj}
to account for absorption by neutral gas with solar abundances ({\tt
abs} model within SPEX, and {\tt wabs} model in XSPEC).  Spectral
uncertainties are given at 90\% confidence ($\Delta$\chisq = 2.71 for
one interesting parameter), and upper limits at 95\% confidence. The
\chisq\ values obtained from SPEX were calculated using the expected
errors based on the model rather than the observed errors (see the
SPEX user's manual). We quote all $EW$s with positive values for both
absorption and emission features. We rebinned the EPIC pn spectra to
over-sample the full-width at half-maximum ($FWHM$) of the energy
resolution by a factor 3, and to have a minimum of 25 counts per bin
to allow the use of the $\chi^2$ statistic. To account for systematic
effects we added quadratically a 2\% uncertainty to each pn spectral
bin.

In order to study the different sources in a systematic manner, we
used for all the spectra the same continuum consisting of blackbody
and power-law components modified by neutral absorption (the {\tt
abs*(bb+pl)} model in SPEX). For each source we fit simultaneously all
the EPIC pn spectra of the persistent and dipping intervals with the
continuum parameters tied together, while we allowed to vary all the
other parameters.  We included Gaussian emission profiles when
emission features were evident near 1~keV and/or 6~keV.  The 1~keV
feature has been previously modeled either as an emission line, or as
an edge, and its nature is unclear \citep[e.g.,][]{1658:sidoli01aa,
1254:boirin03aa, 1916:boirin04aa, 1323:boirin05aa}. Examination of the
spectral residuals reveals strong absorption features around 7~keV for
all the sources except \exo\ and \fouru. To account for these narrow
features, we included absorption from a photo-ionized plasma ({\tt
xabs} in SPEX) in the spectral model.

The \xabs\ model of SPEX treats, in a simplified manner, the
absorption by a thin slab composed of different ions, located between
the ionizing source and the observer. It assumes that the angle
subtended by the slab as seen from the ionizing source is
small. Therefore, emission from the slab and scattering by the slab of
the ionizing source into the line-of-sight are neglected, and only
absorption and scattering out of the line-of-sight by the slab are
considered. The processes taken into account are the continuum and the
line absorption by the ions and scattering out of the line-of-sight by
the free electrons in the slab. Most continuum opacities are taken
from \citet{verner95aa}, while line opacities and wavelengths for most
ions are taken from \citet{verner96ad} (see the details and additional
references in the SPEX user's manual).  The relative column densities
of the ions are coupled through a photo-ionization model. Using codes
such as CLOUDY \citep{ferland03araa}, and assuming a broad-band
ionizing continuum from infra-red to hard X-rays, the ionic column
densities of a photo-ionized slab can be pre-calculated for different
values of $\xi$. SPEX reads in the grid of pre-calculated ionic column
densities during the fitting process. In this paper, we use a grid of
ionic column densities pre-calculated using CLOUDY and assuming that
the ionizing continuum for each source may be represented by a cutoff
power-law ($E^{-\Gamma} \exp(-E/E{\rm _c})$) with a power-law index,
\phind, obtained from the fits to the EPIC pn persistent spectra with
the cutoff energy, $E_c$, fixed to the value measured with BeppoSAX or
R-XTE (see Table~\ref{tab:ioniz-cont}).

The parameters of the \xabs\ model that were allowed to vary are
\nhxabs, $\xi$ and \sigmav. \nhxabs\ is the equivalent hydrogen column
density of the ionized absorber in units of atom~cm$^{-2}$.  $\xi$ is
the ionization parameter of the absorber, defined as $\xi = L /n_{\rm
e} \, r^{2}$, where $L$ is the luminosity of the ionizing source,
$n{\rm _e}$ the electron density of the plasma and $r$ the distance
between the slab and the ionizing source. $\xi$ is expressed in units
of \xiunit, but we will omit the units when quoting \logxi\ values in
this paper. \sigmav\ is the turbulent velocity broadening of the
absorber in \kms, defined as $\sigma_{\rm total}^{2} = \sigma_{\rm v
}^{2} + \sigma_{\rm thermal}^{2}$, where $\sigma_{\rm total}$ is the
total width of a line and $\sigma_{\rm thermal}$ the thermal
contribution.

We present now the results of the
fits for each of the sources individually before discussing the
general properties of the absorbers in dipping LMXBs
 in Sect.~\ref{sec:phenomenology}.


\begin{table*}
\caption{\nineteen: best-fits to the EPIC pn persistent and 5 dip
spectra using the {\tt abs*xabs*(pl+bb)+abs*(gau)} model. The
components of the continuum {\tt (pl+bb)} are identical for all
the spectra and only the \nh\ of the neutral absorber, the
emission line parameters and \nhxabs, $\sigma _v$ and $\xi$ of the
ionized absorber are individually fit for each spectrum. \flux\ is
the 0.6--10 keV absorbed flux. The $FWHM$ of the Gaussian emission line
is constrained to be $\le$0.2~keV.}\begin{tabular}{lccccccc}
\hline \hline\noalign{\smallskip}
 & & Persistent & Dip 1 & Dip 2 & Dip 3 & Dip 4 & Dip 5 \\
\noalign{\smallskip\hrule\smallskip}
& Comp. & & & & &  &  \\
Parameter & & & & & &  & \\
& {\tt pl} & & & & & & \\
\phind & & \multicolumn{6}{c}{2.25 $\pm$ 0.03}  \\
\multicolumn{2}{l}{\kpl\ {\small ($10^{44}$ ph. s$^{-1}$ keV$^{-1}$)}} & \multicolumn{6}{c}{10.0 $\pm$ 0.2} \\
& {\tt bb} & & & & & &\\
\ktbb\ {\small(keV)} & & \multicolumn{6}{c}{1.95 $\pm$ 0.04}  \\
\kbb\  {\small($10^{11}$ cm$^{2}$)} & & \multicolumn{6}{c}{0.65 $\pm$ 0.08} \\
& {\tt abs} & & & & & &  \\
\multicolumn{2}{l}{\nhabs\ {\small($10^{22}$ cm$^{-2}$)}} & 0.432 $\pm$ 0.002 & 0.444 $\pm$ 0.004 & 0.450 $\pm$ 0.006 & 0.54 $\pm$ 0.02 & 0.60 $\pm$ 0.02 & 0.89 $\pm$ 0.07 \\
& {\tt xabs} & & & & & & \\
\multicolumn{2}{l}{\nhxabs\ {\small($10^{22}$ cm$^{-2}$)}} & 4.2 $\pm$ 0.5 & 9 $\pm$ 2 & 16 $\pm$ 2 & 22 $\pm$ 2 & 28 $\pm$ 2  & 54 $\pm$ 3  \\
\logxi\ {\small(\xiunit)} & & 3.05 $\pm$ 0.04 & 2.84 $\pm$ 0.04 & 2.75 $\pm$ 0.02 & 2.62$\,^{+0.02}_{-0.06}$ & 2.55 $\pm$ 0.04 & 2.52$\,^{+0.02}_{-0.06}$ \\
\sigmav\ {\small(km s$^{-1}$)} & & 2300$\,^{+2100}_{-1700}$ & 50 $\pm$ 30 & 90 $\pm$ 25 & 120 $\,^{+40}_{-20}$ & 220$\,^{+80}_{-90}$ & 160$\,^{+100}_{-40}$ \\
\noalign {\smallskip}
& {\tt gau}  & & & & & & \\
\multicolumn{2}{l}{ \egau\ {\small(keV)}}                &  &  &  & 0.89 $\pm$ 0.03 & 0.92 $\pm$ 0.04 & 0.88 $\pm$ 0.04 \\
\multicolumn{2}{l}{ $FWHM$ {\small(keV)}}               &  &  &   & 0.2 & 0.2 & 0.2 \\
\multicolumn{2}{l}{ \kgau\ {\small(10$^{44}$ ph s$^{-1}$)}} &  &  &  & 0.6 $\pm$ 0.1 & 0.6 $\pm$ 0.1 & 0.5 $\pm$ 0.3 \\
\noalign {\smallskip}
\hline\noalign {\smallskip}
\multicolumn{2}{l}{\flux\ \small (10$^{-10}$ \ergcms)}  & 2.6 & 2.4 & 2.0 & 1.5 & 1.2 & 0.7 \\
\multicolumn{2}{l}{\rchisq (d.o.f.)} & \multicolumn{6}{c}{1.30 (1264)} \\

        \multicolumn{2}{l}{Exposure (ks)} & 9.1 & 2.1 & 1.1 & 0.9 & 0.6 & 0.8 \\
\noalign{\smallskip\hrule\smallskip}
\label{tab:bestfit-1916}
\end{tabular}
 
\label{tab:1916-bestfit}
\end{table*}

\begin{figure*}[ht!]
\centerline{\includegraphics[angle=0,width=0.47\textwidth]{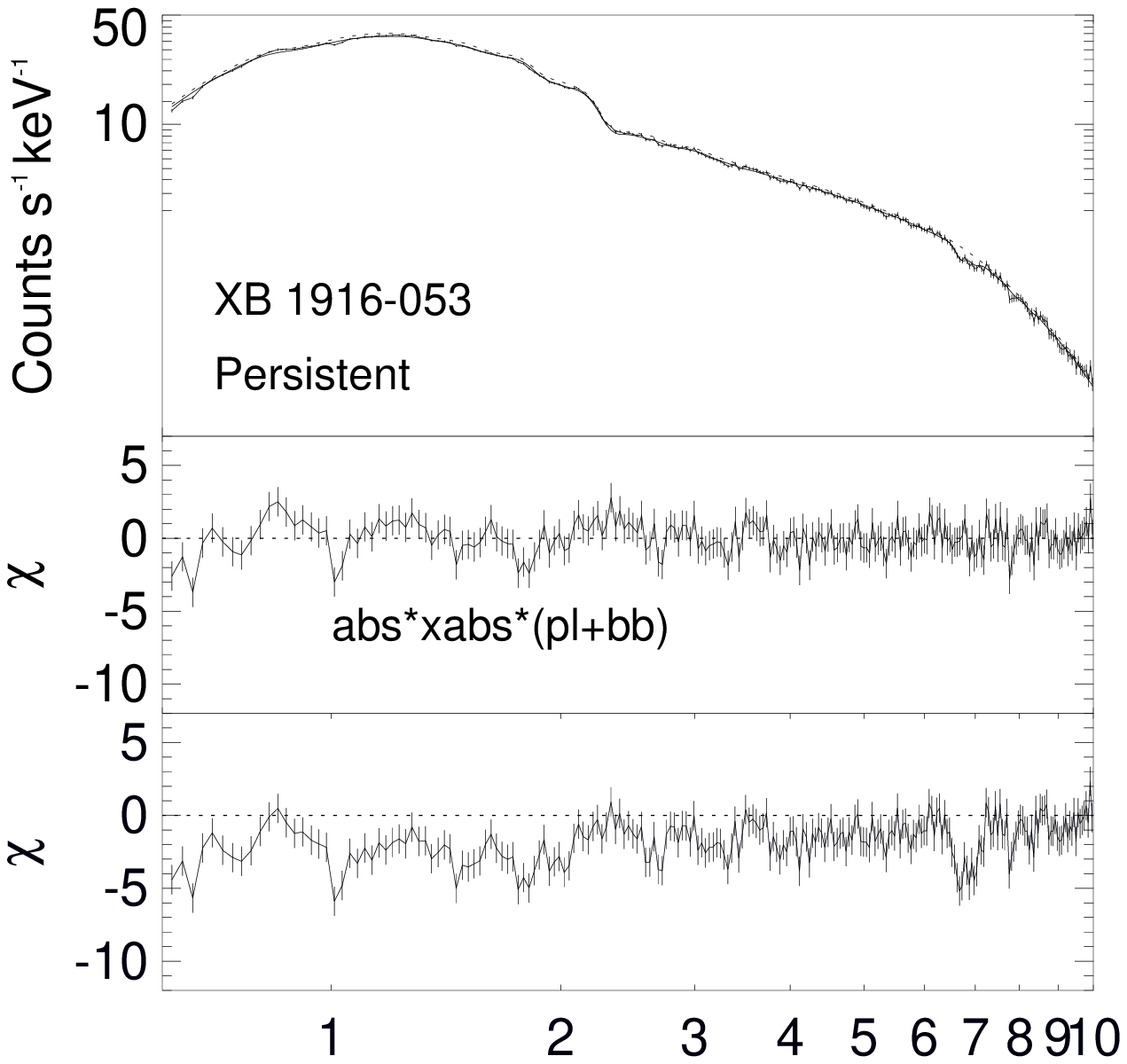}
\hspace{-2.8cm}
\includegraphics[width=0.47\textwidth]{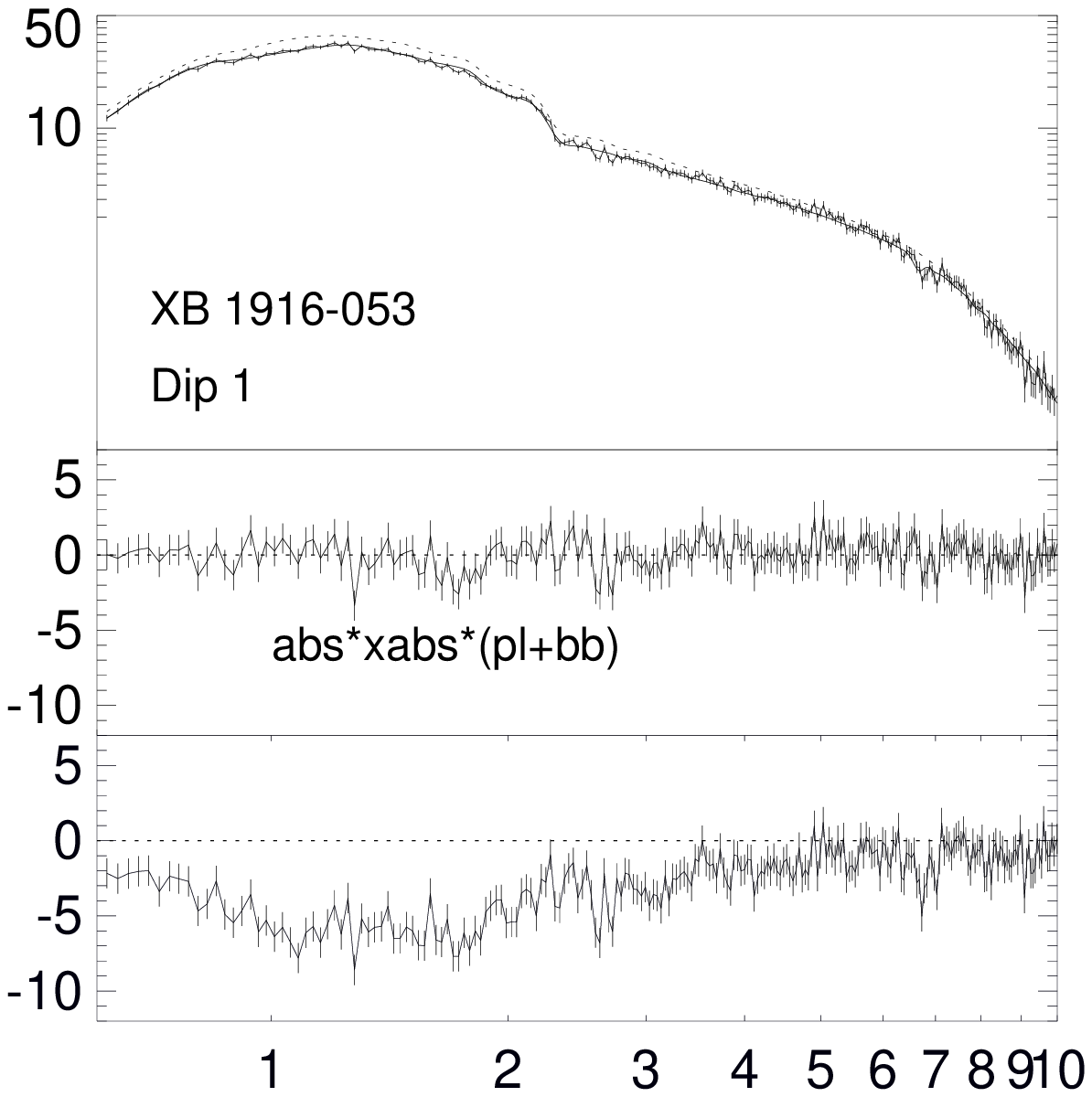}
\hspace{-2.8cm}
\includegraphics[width=0.47\textwidth]{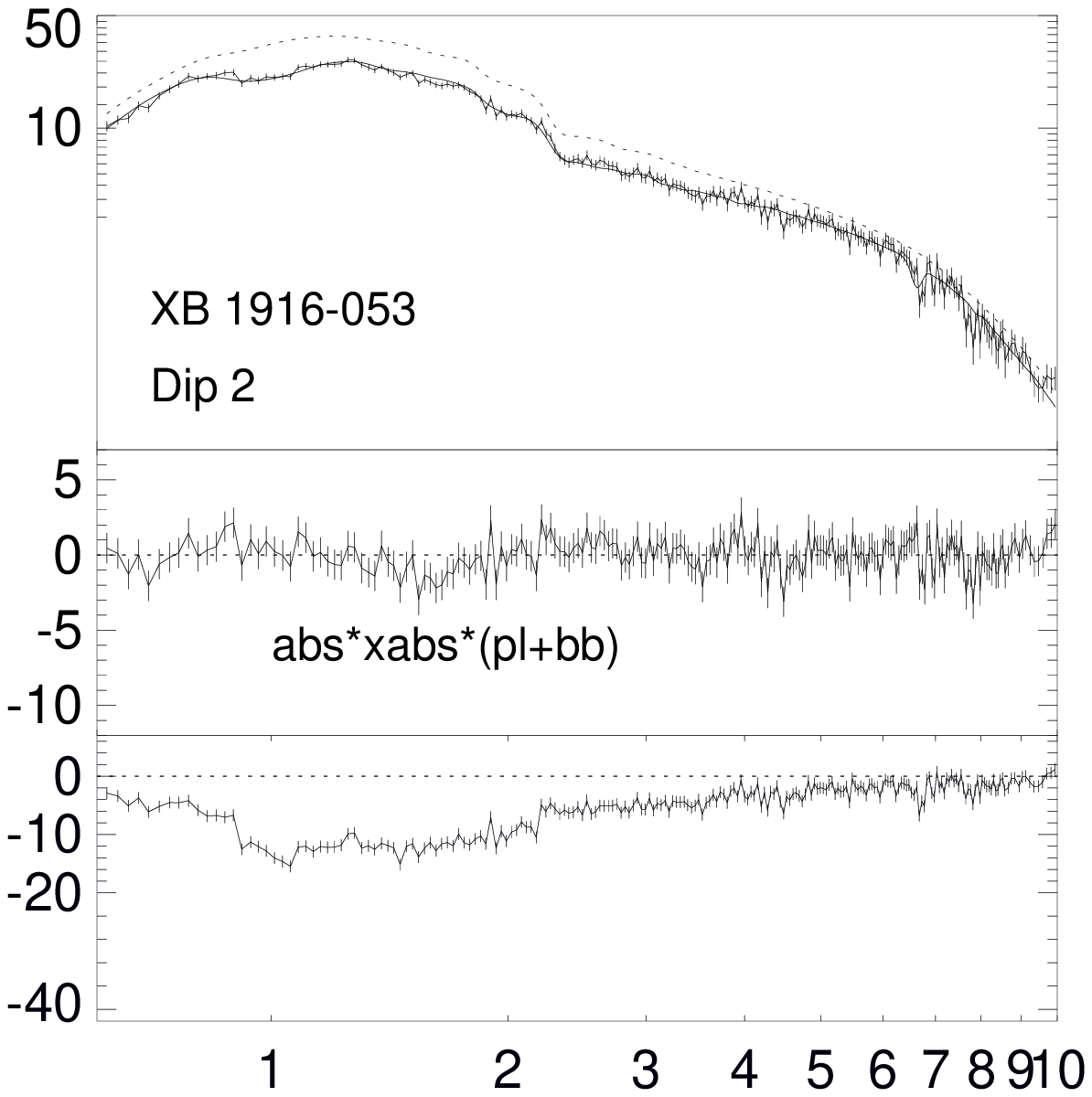}}
\vspace{-0.2cm} \centerline{
\includegraphics[angle=0,width=0.47\textwidth]{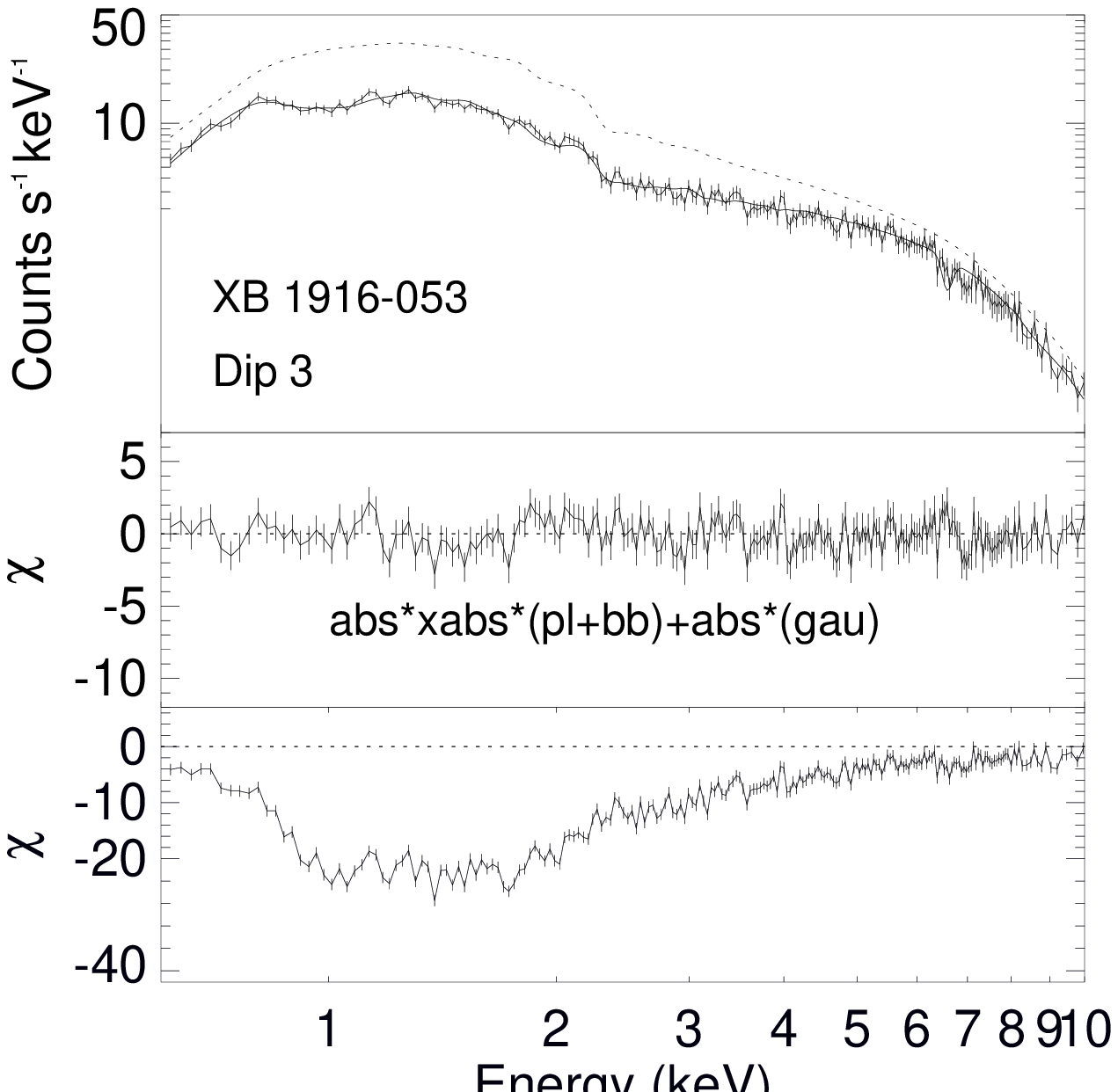}
\hspace{-2.8cm}
\includegraphics[width=0.47\textwidth]{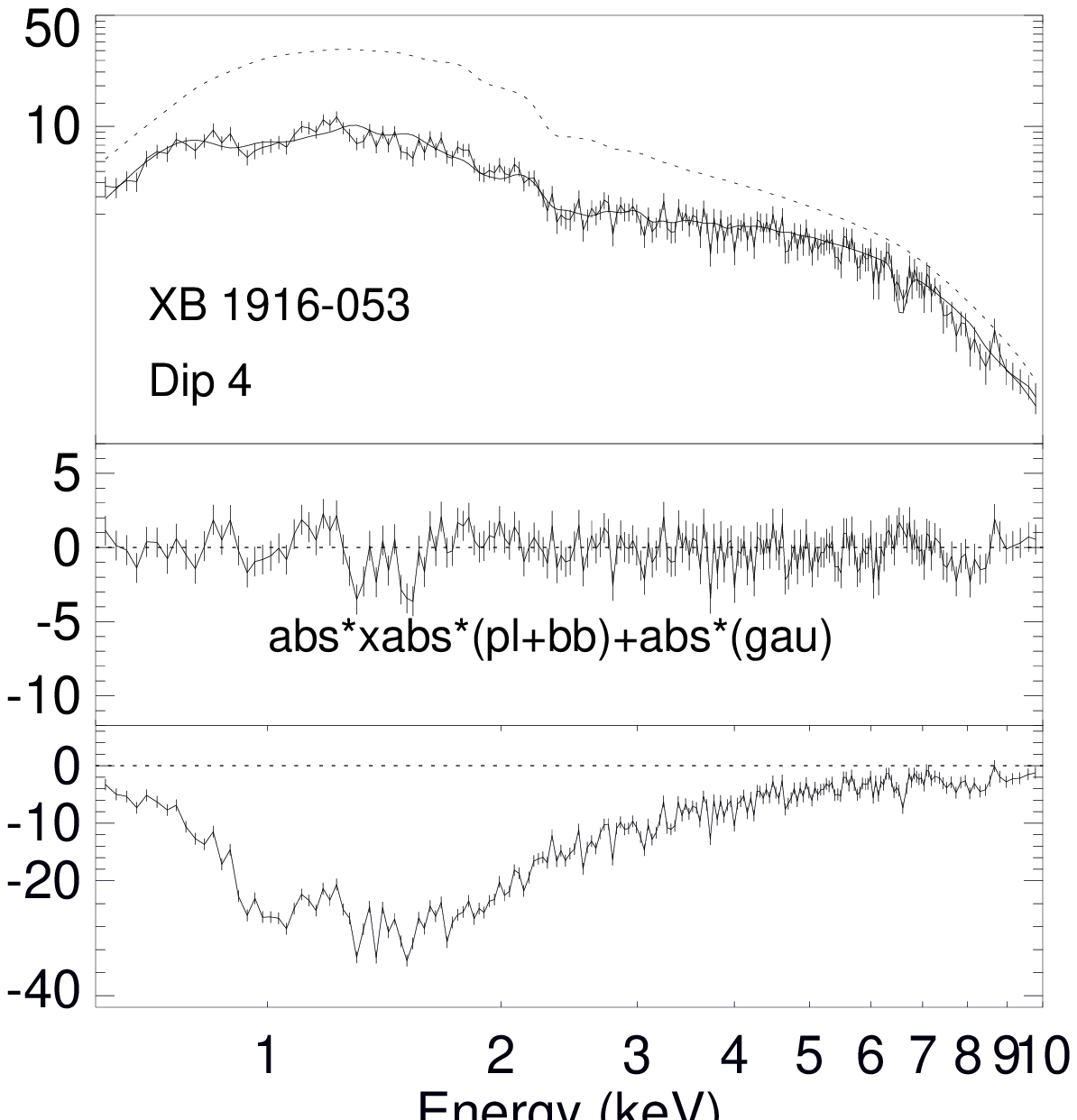}
\hspace{-2.8cm}
\includegraphics[width=0.47\textwidth]{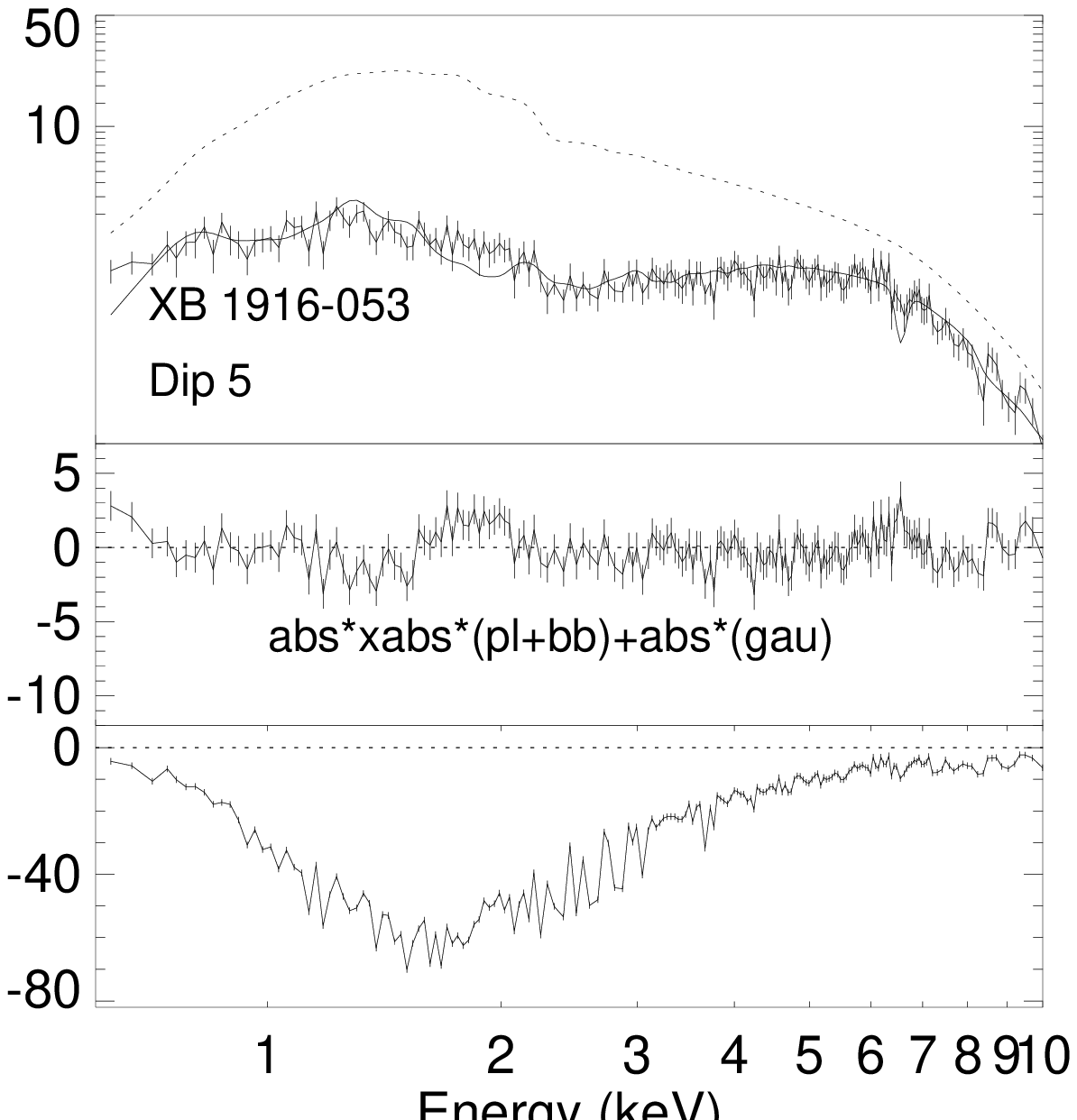}
\vspace{0.2cm} } \caption{\nineteen: EPIC pn persistent and dipping
spectra fit with a power-law ({\tt pl}) and a blackbody ({\tt bb})
model, modified by absorption from neutral ({\tt abs}) and ionized
({\tt xabs}) material plus one narrow emission line ({\tt gau}),
modified by absorption from neutral material ({\tt abs}) (see
Table~\ref{tab:bestfit-1916}).  The dotted lines show the models when
the column of the ionized absorber is set to 0.  The middle panels
show residuals in units of standard deviations from the above
model. The lower panels show residuals when \nhxabs\ is set to 0.  }
\label{fig:bestfit-1916-2}
\end{figure*}

\subsection{\nineteen}

Results of the fits of the \nineteen\ persistent and dipping spectra
to the {\tt abs*xabs*(pl+bb) + abs*(gau)} SPEX model are shown in
Fig.~\ref{fig:bestfit-1916-2} and the best-fit parameter values are
listed in Table~\ref{tab:bestfit-1916}.  The line and continuum
changes can be primarily modeled by changes in the ionized absorber
with \nhxabs\ increasing from $(4.2 \pm 0.5) \times 10^{22}$ to $(54
\pm 3) \times 10^{22}$~atom~cm$^{-2}$ while \logxi\ decreases from
$3.05 \pm 0.04$ to $2.52\,^{+0.02}_{-0.06}$, with a smaller change in
the amount of neutral absorption, \nh, which increases from $(0.432
\pm 0.002) \times 10^{22}$ to $(0.89 \pm 0.07) \times
10^{22}$~atom~cm$^{-2}$. The $\sim$1~keV feature is only visible in
the Dip 3 to Dip 5 spectra (see Table \ref{tab:dipselection} for Dip
definition).  The strongest absorption lines in the persistent
spectrum are from \fetfive\ and \fetsix\ with $EW$s of 34~eV and
21~eV, respectively.  The \fetfive\ and \fetsix\ \lyb\ features are
also predicted by the \xabs\ model with $EW$s of 7~eV and 4~eV,
respectively. During dipping, the \ew\ of the \fetsix\ line decreases
and a feature is observed at an energy 6.6 $\pm$ 0.1~keV, possibly
originating from a blend of \fetfive\ to \fenineteen\ lines.  The
best-fit value of \logxi\ is $3.05 \pm 0.04$ for the persistent
spectrum. This is significantly lower than the value of 3.92 reported
by \citet[]{1916:boirin04aa}. This difference may be explained by the
fact that their result is based on the photo-ionized model by
\citet{kallman01apjs} which assumes an ionizing continuum consisting
of a power-law with \phind\ = 1 while we use as ionizing continuum a
cutoff power-law of \phind\ = 1.87 and $E_c$ = 80~keV (see Table
\ref{tab:ion-con}).


\begin{table*}
\caption{\exo: best-fits to the pn persistent and 5 dip spectra using
the {\tt abs*xabs*(pl+bb)+abs*(gau+gau)} model.  The continuum
parameters {\tt (pl+bb)} are the same for all spectra and only the
\nh\ of the neutral absorber, the line parameters, and \nhxabs,
$\sigma _v$ and $\xi$ of the ionized absorber are fit for each
spectrum. \flux\ is the 0.6--10~keV absorbed flux. \nh\ is constrained
to be $\ge$$1.1 \times 10^{21}$~atom~cm$^{-2}$ \citep{0748:sidoli05aa}
and the $FWHM$s of the emission lines are constrained to be
$\le$0.20~keV.}  \begin{center}
\begin{tabular}{lccccccc}
\hline \hline\noalign{\smallskip}
 & & Persistent & Dip 1 & Dip 2 & Dip 3 & Dip 4 & Dip 5 \\
\noalign{\smallskip\hrule\smallskip}
& Comp. & & & & & &  \\
Parameter & & & & & & &  \\
& {\tt po} & & & & & & \\
\phind & & \multicolumn{6}{c}{1.57 $\pm$ 0.05}  \\
\multicolumn{2}{l}{\kpl\ {\small ($10^{44}$ ph. s$^{-1}$ keV$^{-1}$)}} & \multicolumn{6}{c}{3.9 $\pm$ 0.2} \\
& {\tt bb} & & & & & & \\
\ktbb\ {\small(keV)} & & \multicolumn{6}{c}{1.9 $\pm$ 0.1}  \\
\kbb\  {\small($10^{11}$ cm$^{2}$)} & & \multicolumn{6}{c}{0.49 $\pm$ 0.07}  \\
& {\tt abs} & & & & & & \\
\multicolumn{2}{l}{\nh\ {\small($10^{22}$ cm$^{-2}$)}} & 0.11 & 0.11 & 0.11 & 0.21$\,^{+0.02}_{-0.05}$ & 0.27 $\pm$ 0.12   & 0.24$\,^{+0.09}_{-0.05}$   \\
& {\tt xabs} & & & & & &  \\
\multicolumn{2}{l}{\nhxabs\ {\small($10^{22}$ cm$^{-2}$)}} & 3.5 $\pm$ 0.2 & 2.3 $\pm$ 0.3 & 3.3 $\pm$ 0.3 & 4.9 $\pm$ 0.3 & 7.8 $\pm$ 0.4 & 15.5 $\pm$ 0.5 \\
\logxi\ {\small(\xiunit)} & & 2.45 $\pm$ 0.02 & 2.38 $\pm$ 0.05 & 2.23 $\pm$ 0.07 & 2.23 $\pm$ 0.05 & 2.21 $\pm$ 0.04 & 2.26 $\pm$ 0.03 \\
\sigmav\ {\small(km s$^{-1}$)} & &  13 $\pm$ 6 & $<$10 & 26 $\,^{+30}_{-20}$ & 84$\,^{+40}_{-50}$ & 140$\,^{+90}_{-60}$ & 130$\,^{+80}_{-70}$ \\
\noalign {\smallskip}
& {\tt gau}  & & & & & &  \\
\multicolumn{2}{l}{ \egau\ {\small(keV)}} & \multicolumn{6}{c}{0.569 (fixed)} \\
\multicolumn{2}{l}{ $FWHM$ {\small(keV)}} & 0.20 & 0.20 & 0.20 & 0.20 & 0.17 $\pm$ 0.01 & $<$0.15 \\
\multicolumn{2}{l}{ \kgau\ {\small(10$^{44}$ ph s$^{-1}$)}} & 1.85$\,^{+0.14}_{-0.07}$ & 2.3$\,^{+0.3}_{-0.1}$ & 1.6$\,^{+2.1}_{-0.1}$ & 2.9$\,^{+0.5}_{-1.2}$ & 2$\,^{+3}_{-2}$ & 0.4$\,^{+1.1}_{-0.2}$  \\
\noalign {\smallskip}
& {\tt gau}  & & & & & & \\
\multicolumn{2}{l}{ \egau\ {\small(keV)}}               &  \multicolumn{6}{c}{0.915 (fixed)}  \\
\multicolumn{2}{l}{ $FWHM$ {\small(keV)}}               &  0.18 $\pm$ 0.03 & 0.20 
&  0.13$\,^{+0.07}_{-0.05}$ & 0.20 
& 0.11 $\pm$ 0.04 & 0.11 $\pm$ 0.03 \\
\multicolumn{2}{l}{ \kgau\ {\small(10$^{44}$ ph s$^{-1}$)}} & 0.18 $\pm$ 0.03 & 0.17 $\pm$ 0.03 & 0.09$\,^{+0.08}_{-0.03}$ & 0.13 $\pm$ 0.04 & 0.06$\,^{+0.04}_{-0.02}$ & 0.04 $\pm$ 0.02 \\
\noalign {\smallskip}
\noalign {\smallskip}
\hline\noalign {\smallskip}
\multicolumn{2}{l}{\flux\ \small (10$^{-10}$ \ergcms)}  & 2.2 & 2.3 & 2.1 & 1.9 & 1.6 & 1.3 \\
        \multicolumn{2}{l}{\rchisq\ (d.o.f)} & \multicolumn{6}{c}{1.21 (1341)}  \\
        \multicolumn{2}{l}{Exposure (ks)} & 17.6 & 4.4 & 3.6 & 8.2 & 7.1 & 8.6  \\
\noalign{\smallskip\hrule\smallskip}
\label{tab:bestfit-exo}
\end{tabular}
\end{center}
 \label{tab:bestfit-exo}
\end{table*}

\begin{figure*}[ht!]
\centerline{\includegraphics[angle=0,width=0.47\textwidth]{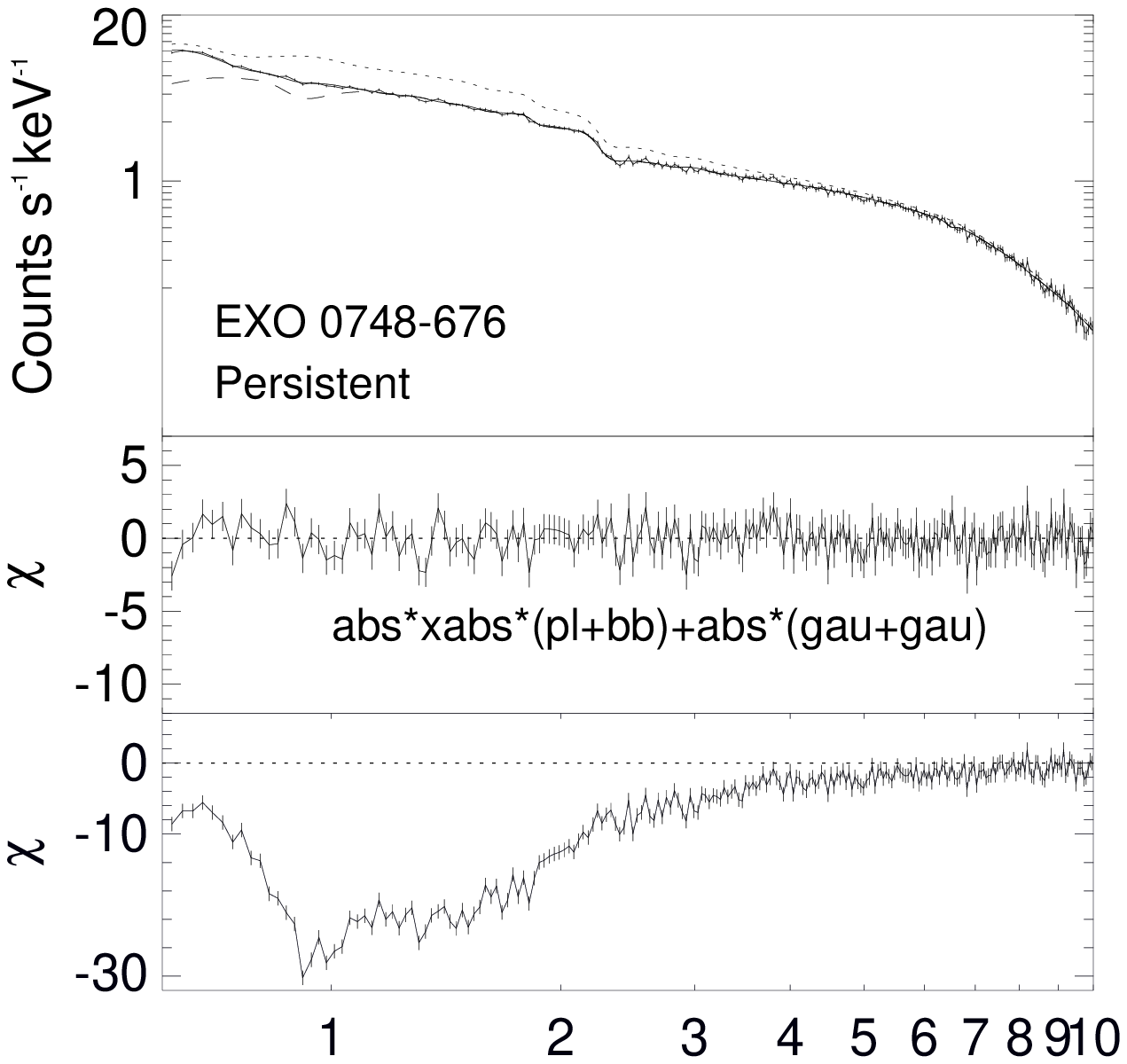}
\hspace{-2.8cm}
\includegraphics[width=0.47\textwidth]{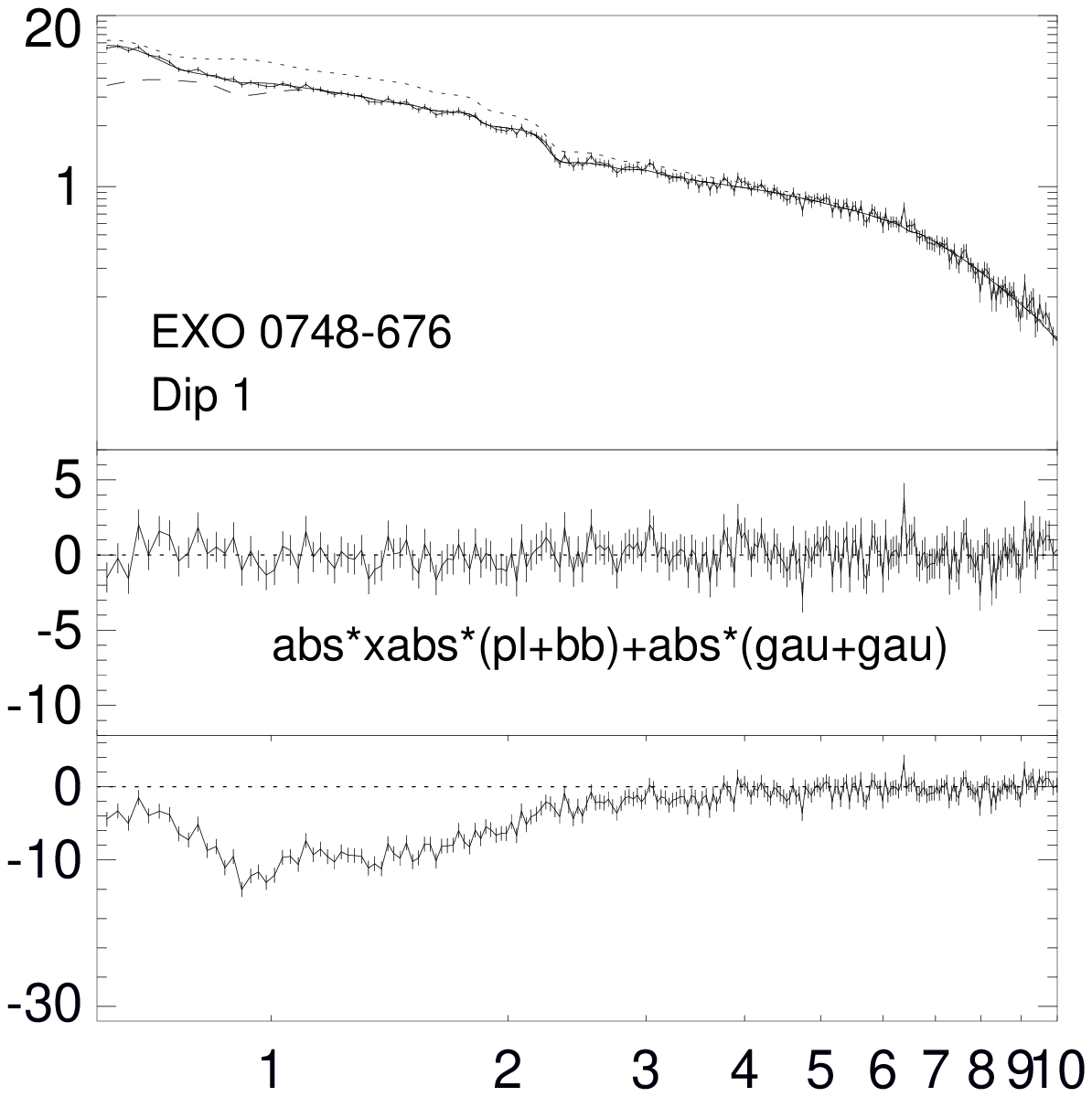}
\hspace{-2.8cm}
\includegraphics[width=0.47\textwidth]{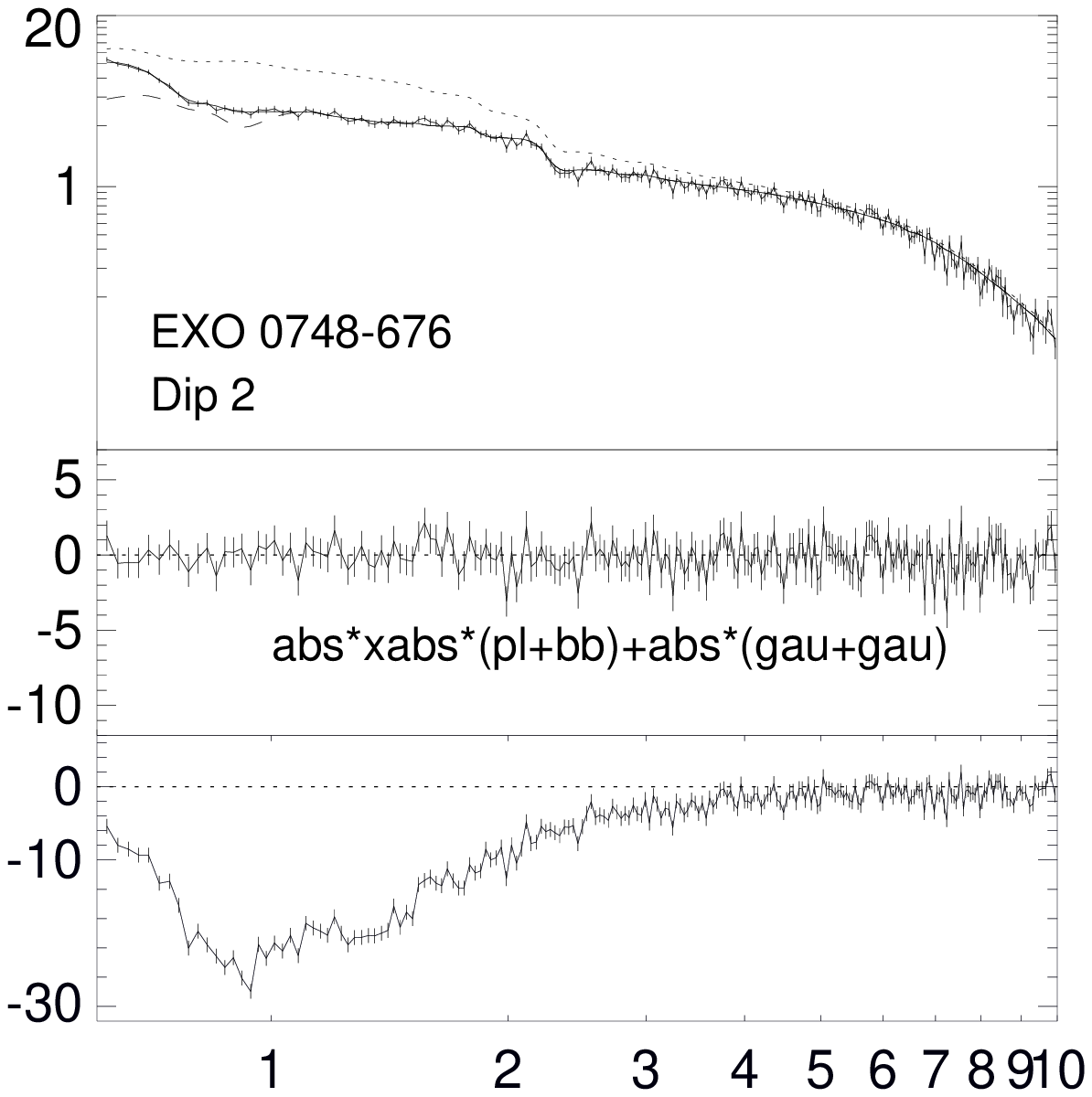}
} \vspace{0.cm}
\centerline{\includegraphics[width=0.47\textwidth]{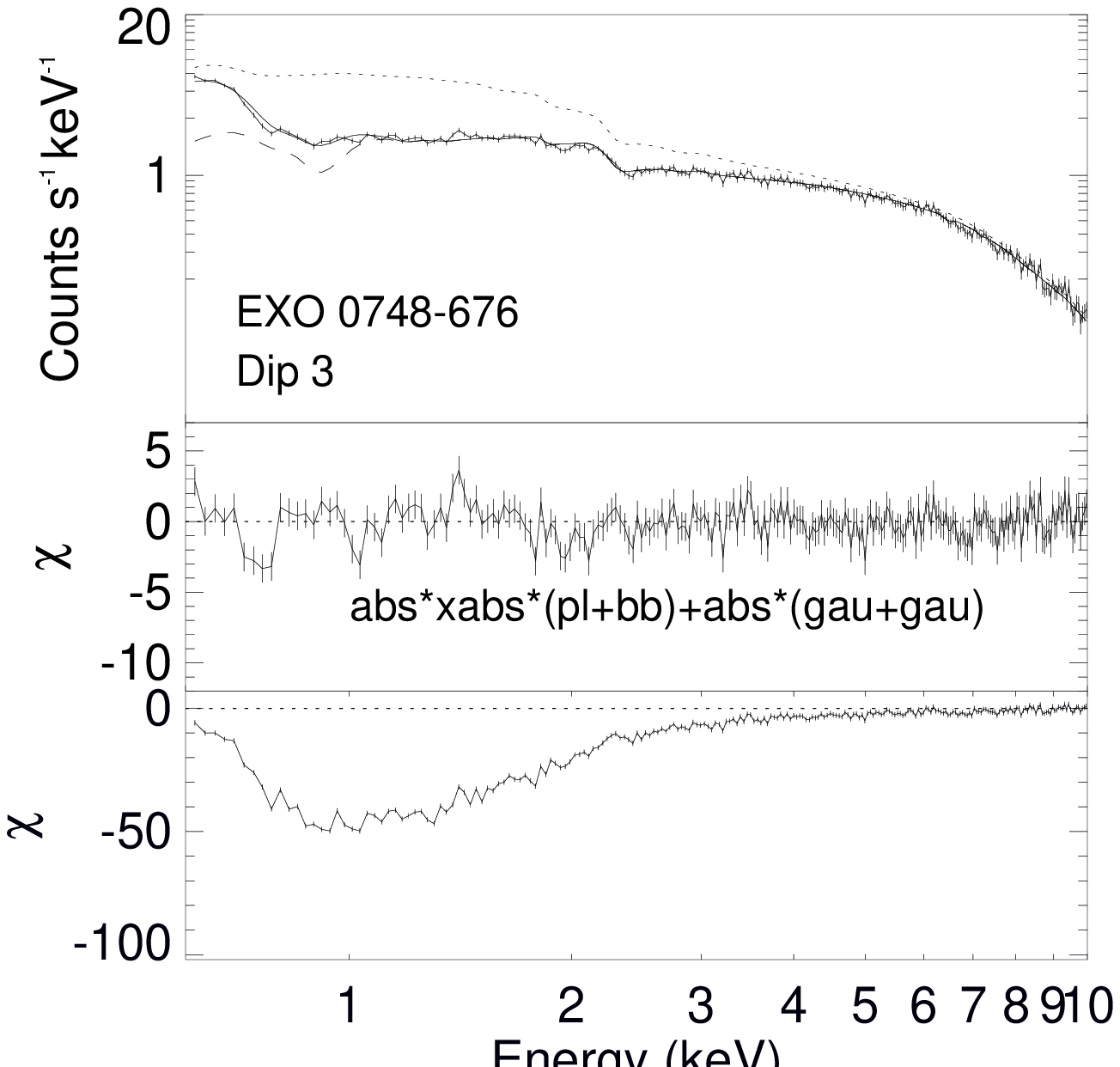}
\hspace{-2.8cm}
\includegraphics[angle=0,width=0.47\textwidth]{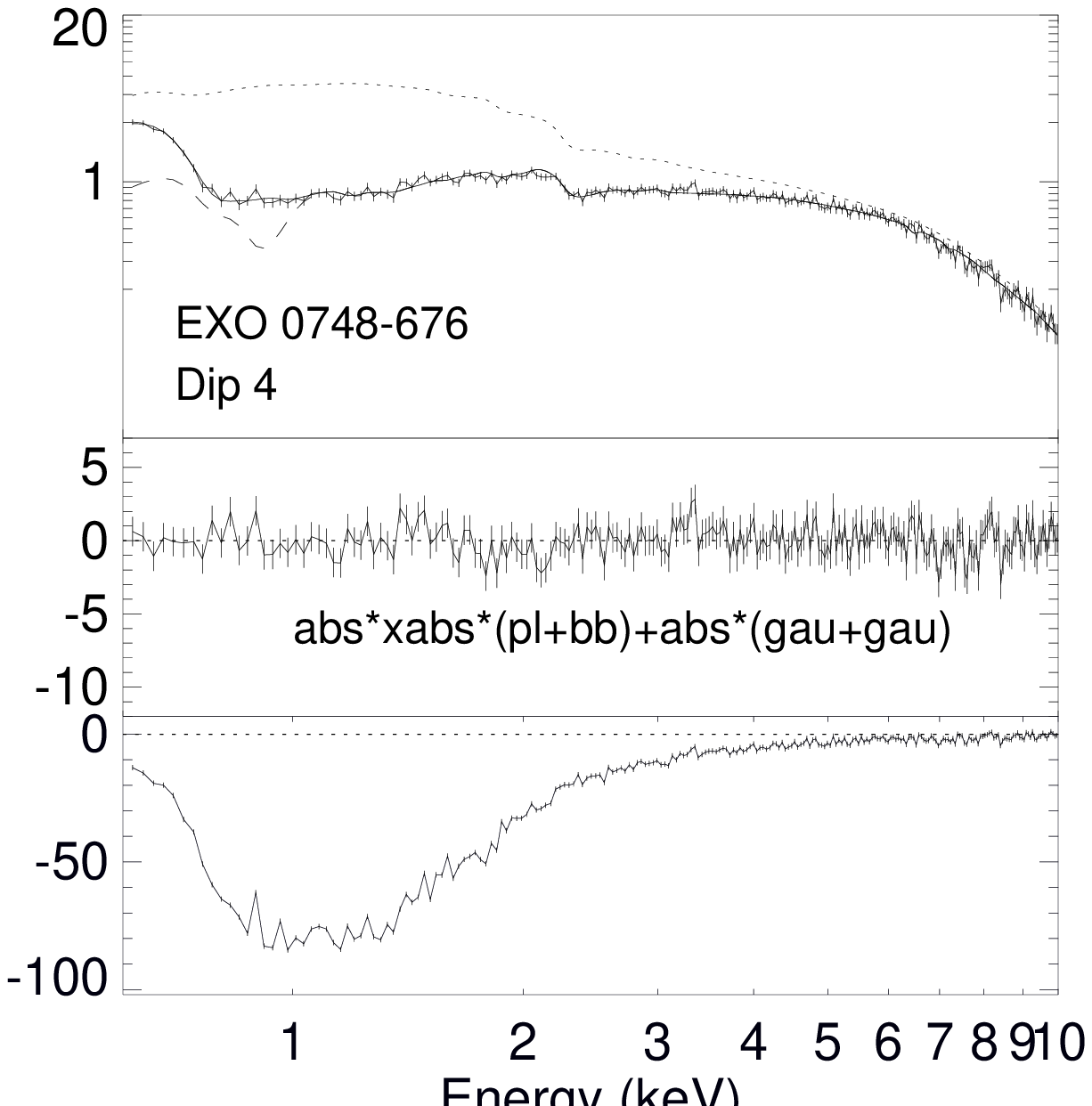}
\hspace{-2.8cm}
\includegraphics[width=0.47\textwidth]{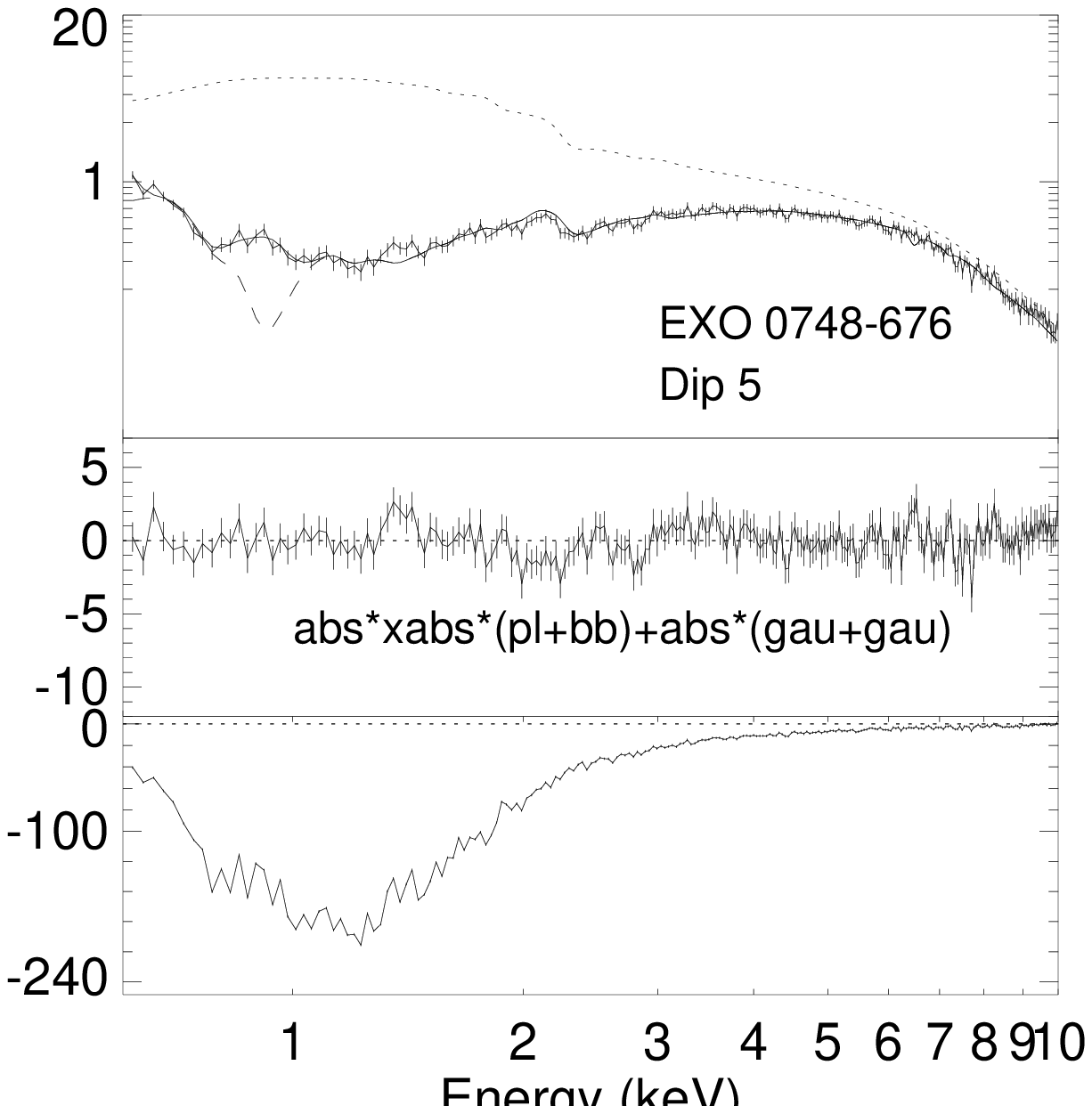}
\vspace{0.2cm}}
\caption{\exo: EPIC pn persistent and 5 dipping spectra fit with a
power-law ({\tt pl}) and blackbody ({\tt bb}) modified by absorption
from neutral ({\tt abs}) and ionized ({\tt xabs}) material with 2
narrow emission lines ({\tt gau}) (see Table~\ref{tab:bestfit-exo}).
The dotted lines show the models when the column of the ionized
absorber is set to 0. The dashed lines show the models when the
normalizations of the Gaussian emission features are set to 0. The
middle panels show the residuals in units of standard deviations from
the above model. The lower panels show the residuals when \nhxabs\ is
set to 0.} \label{fig:bestfit-exo}
\end{figure*}

\subsection{\exo}
\label{sec:exo}

Narrow X-ray absorption lines have been detected from
\exo\ during X-ray bursts \citep{0748:cottam02nat}, but such
features have not been detected during persistent or dipping intervals.
The 0.6--10~keV pn persistent spectrum was
first fit using a blackbody and a power-law model, together
with photo-electric absorption by neutral material.
The fit is unacceptable with a
\rchisq\ of $\sim$10 for 235 degrees of freedom (d.o.f.) due
mainly to the presence of strong soft excesses at 2~keV and $\approxlt$1~keV.
Even though no strong narrow absorption lines are evident,
we investigated the possibility that the soft excesses result
from the presence of an ionized absorber by fitting the persistent
spectrum with the standard {\tt abs*xabs*(pl+bb)} SPEX model. A
minimum value of $1.1 \times 10^{21}$~atom~cm$^{-2}$ \citep{0748:sidoli05aa}
was imposed on \nh\ to prevent this parameter going to
unrealistically low values. The
soft excesses were both well modeled by the ionized absorber, but
the fit is still formally unacceptable with a \rchisq\ of $\sim$3
for 232 d.o.f. Examination of the residuals reveals narrow
emission features at energies $\approxlt$2~keV, which were modeled by
including two Gaussian emission lines at $0.57$, and $0.92$~keV,
the energies of inter-combination lines of \oseven\ and \nenine,
respectively. The fit is now significantly improved with a
\rchisq\ of 1.4 for 225 d.o.f. Next we fit the persistent and dipping
spectra together in the usual manner with the model {\tt
abs*xabs*(pl+bb)+abs*(gau+gau)}.
The best-fit parameters are given in
Table~\ref{tab:bestfit-exo} and the spectra are shown in
Fig.~\ref{fig:bestfit-exo}. The changes in the continuum can be
modeled primarily by changes in the ionized absorber with a large
increase in \nhxabs\ from (3.5 $\pm$ 0.2$) \times
10^{22}$~atom~cm$^{-2}$ to (15.5 $\pm$ 0.5$) \times
10^{22}$~atom~cm$^{-2}$ while \logxi\ decreases from 2.45 $\pm$
0.02 to 2.26 $\pm$ 0.03. A smaller increase in the
amount of neutral absorption from $0.11 \times
10^{22}$ to (0.24$\,^{+0.09}_{-0.05}) \times 10^{22}$~atom~cm$^{-2}$
is required.
The ionized absorber is not well constrained due to the
absence of strong individually resolved absorption
features in the spectra. In contrast, strong absorption edges are evident,
which are produced e.g., in the persistent emission, by \oeight\
and \neten\ at 0.87 and 1.36~keV with optical depths of 0.39 and 0.09,
respectively.


\begin{table*}[ht!]
\caption{\twelve: best-fits to the EPIC pn persistent and 3 dip
spectra using the {\tt abs*xabs*(pl+bb)+abs*(gau+gau)} model. The
parameters for the continuum components {\tt (pl+bb)} are the same for
all the spectra and only the \nh\ of the neutral absorber, the line
parameters, and \nhxabs, $\sigma _v$ and $\xi$ of the ionized absorber
are individually fit for each spectrum. \flux\ is the 0.6--10 keV
absorbed flux. The $FWHM$ of the 6.8~keV Gaussian emission line is
constrained to be $\le$2.0~keV.  The $FWHM$s of the $\sim$1~keV
Gaussian emission lines are constrained to be $\le$0.2~keV. } \begin{center}
\begin{tabular}{lccccc}
\hline \hline\noalign{\smallskip}
 & & Persistent & Dip 1 & Dip 2 & Dip 3 \\
\noalign{\smallskip\hrule\smallskip}
& Comp. & & & &  \\
Parameter & & & & & \\
& {\tt pl} & & & &  \\
\phind & & \multicolumn{4}{c}{2.09 $\pm$ 0.02}  \\
\multicolumn{2}{l}{\kpl\ {\small ($10^{44}$ ph. s$^{-1}$ keV$^{-1}$)}} & \multicolumn{4}{c}{19.4$\,^{+0.8}_{-0.5}$} \\
& {\tt bb} & & & &\\
\ktbb\ {\small(keV)} & & \multicolumn{4}{c}{1.09 $\pm$ 0.02}  \\
\kbb\  {\small($10^{11}$ cm$^{2}$)} & & \multicolumn{4}{c}{16.1 $\pm$ 0.7}  \\
& {\tt abs} & & & & \\
\multicolumn{2}{l}{\nh\ {\small($10^{22}$ cm$^{-2}$)}} & 0.346 $\pm$ 0.002 & 0.340 $\pm$ 0.003 & 0.350 $\pm$ 0.006 & 0.39 $\pm$ 0.01 \\
& {\tt xabs} & & & &  \\
\multicolumn{2}{l}{\nhxabs\ {\small($10^{22}$ cm$^{-2}$)}} & 8.4 $\pm$ 0.3 & 18.6 $\pm$ 0.8 & 25 $\pm$ 2  & 47 $\pm$ 3 \\
\logxi\ {\small(\xiunit)} & & 4.3 $\pm$ 0.1 & 3.9 $\pm$ 0.3 & 2.98$\,^{+0.06}_{-0.03}$ & 2.94 $\pm$ 0.05 \\
\sigmav\ {\small(km s$^{-1}$)} & & 2800 $\pm$ 1900  & 70 $\,^{+70}_{-50}$ & 130 $\pm$ 35 & 350$\,^{+110}_{-80}$ \\
\noalign {\smallskip}
& {\tt gau}  & & & & \\
\multicolumn{2}{l}{ \egau\ {\small(keV)}}               & 6.8 $\pm$ 0.1                 &                &   &  \\
\multicolumn{2}{l}{ $FWHM$ {\small(keV)}}               &  2.0     &       &   &  \\
\multicolumn{2}{l}{ \kgau\ {\small(10$^{44}$ ph s$^{-1}$)}} &  0.08 $\pm$ 0.01 &         &   & \\
\noalign {\smallskip}
& {\tt gau}  & & & & \\
\multicolumn{2}{l}{ \egau\ {\small(keV)}}               & 1.00 $\pm$ 0.02             &  0.99 $\pm$ 0.04                 &  1.07 $\pm$ 0.03 & 1.06 $\pm$ 0.05 \\
\multicolumn{2}{l}{ $FWHM$ {\small(keV)}}               &  0.2  &  0.2  & $<$0.2 & $<$0.2 \\
\multicolumn{2}{l}{ \kgau\ {\small(10$^{44}$ ph s$^{-1}$)}} & 0.28 $\pm$ 0.06 & 0.22 $\pm$ 0.07         & 0.2$\,^{+0.2}_{-0.1}$ & 0.2$\,^{+0.2}_{-0.1}$ \\
\noalign {\smallskip}
\hline\noalign {\smallskip}
\multicolumn{2}{l}{\flux\ \small (10$^{-10}$ \ergcms) } & 6.4 & 5.8 & 4.8 & 3.4 \\
        \multicolumn{2}{l}{\rchisq (d.o.f.)} & \multicolumn{4}{c}{1.23 (846)}  \\
        \multicolumn{2}{l}{Exposure (ks)} & 11.5 & 2.7 & 0.6 & 0.2 \\
\noalign{\smallskip\hrule\smallskip}
\label{tab:bestfit-1254}
\end{tabular}
\end{center}

\end{table*}

\begin{figure*}[ht!]
\centerline{\includegraphics[angle=0,width=0.47\textwidth]{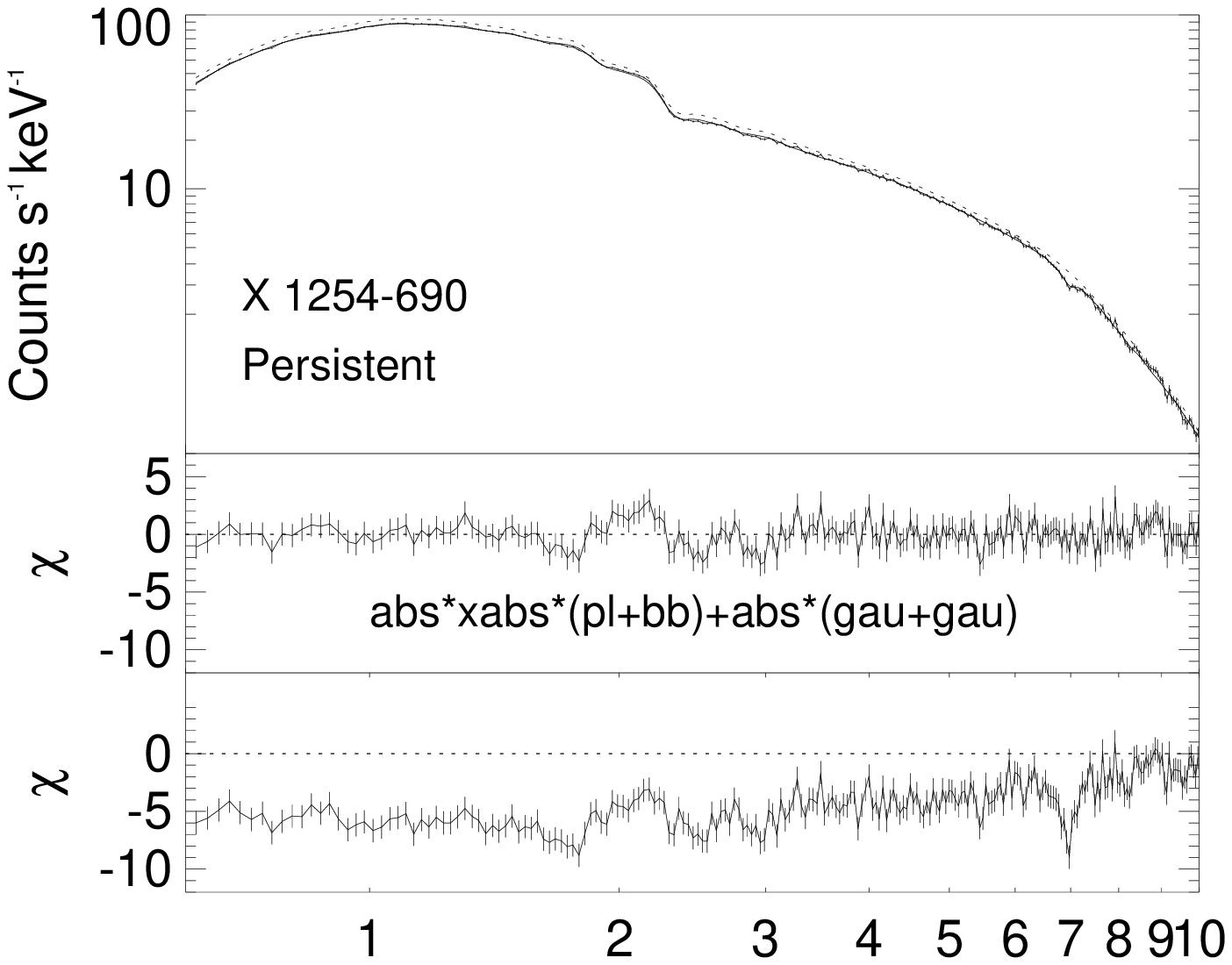}
\hspace{-2.3cm}
\includegraphics[width=0.47\textwidth]{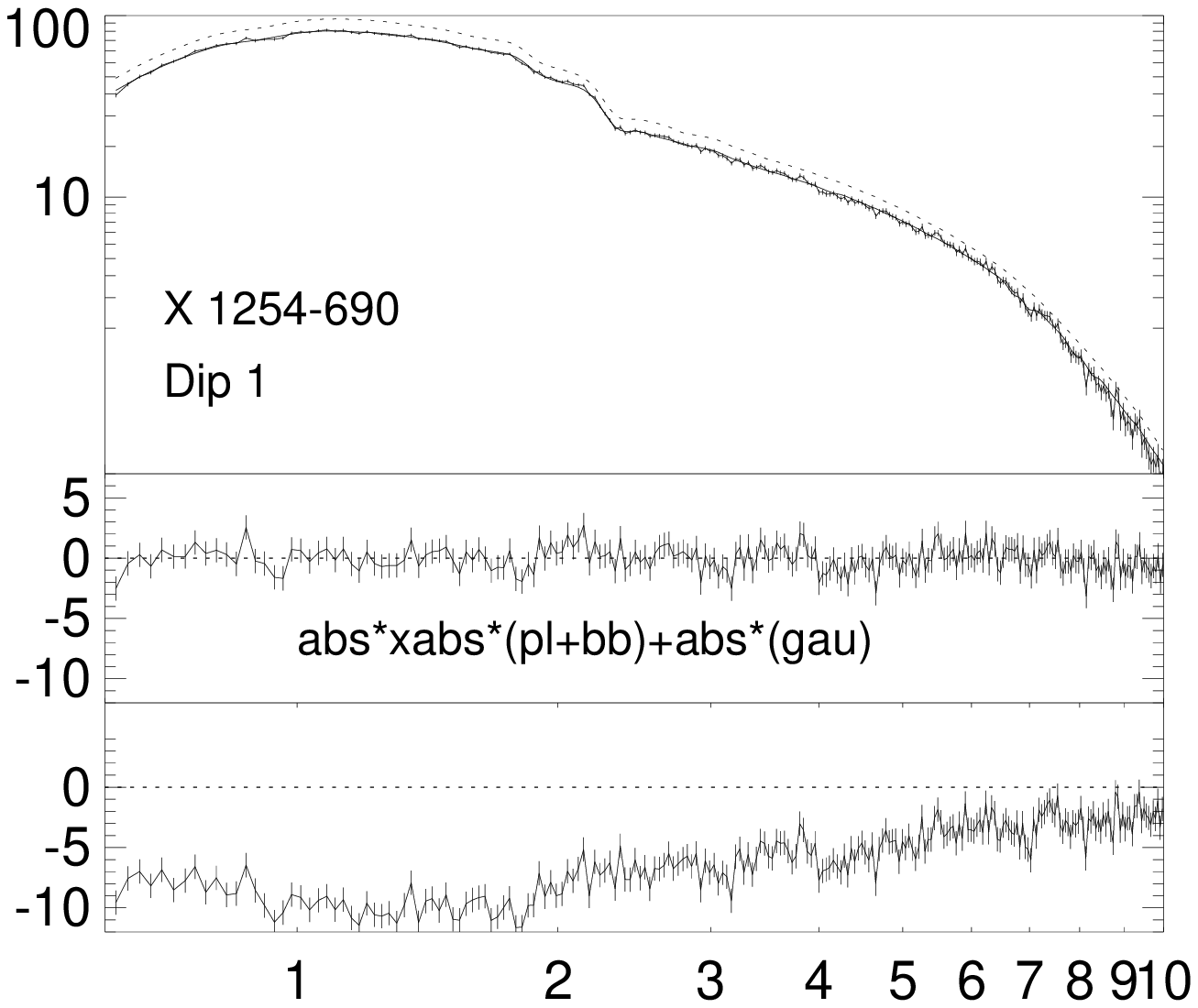}
} \vspace{-0.5cm} \centerline{
\includegraphics[angle=0,width=0.47\textwidth]{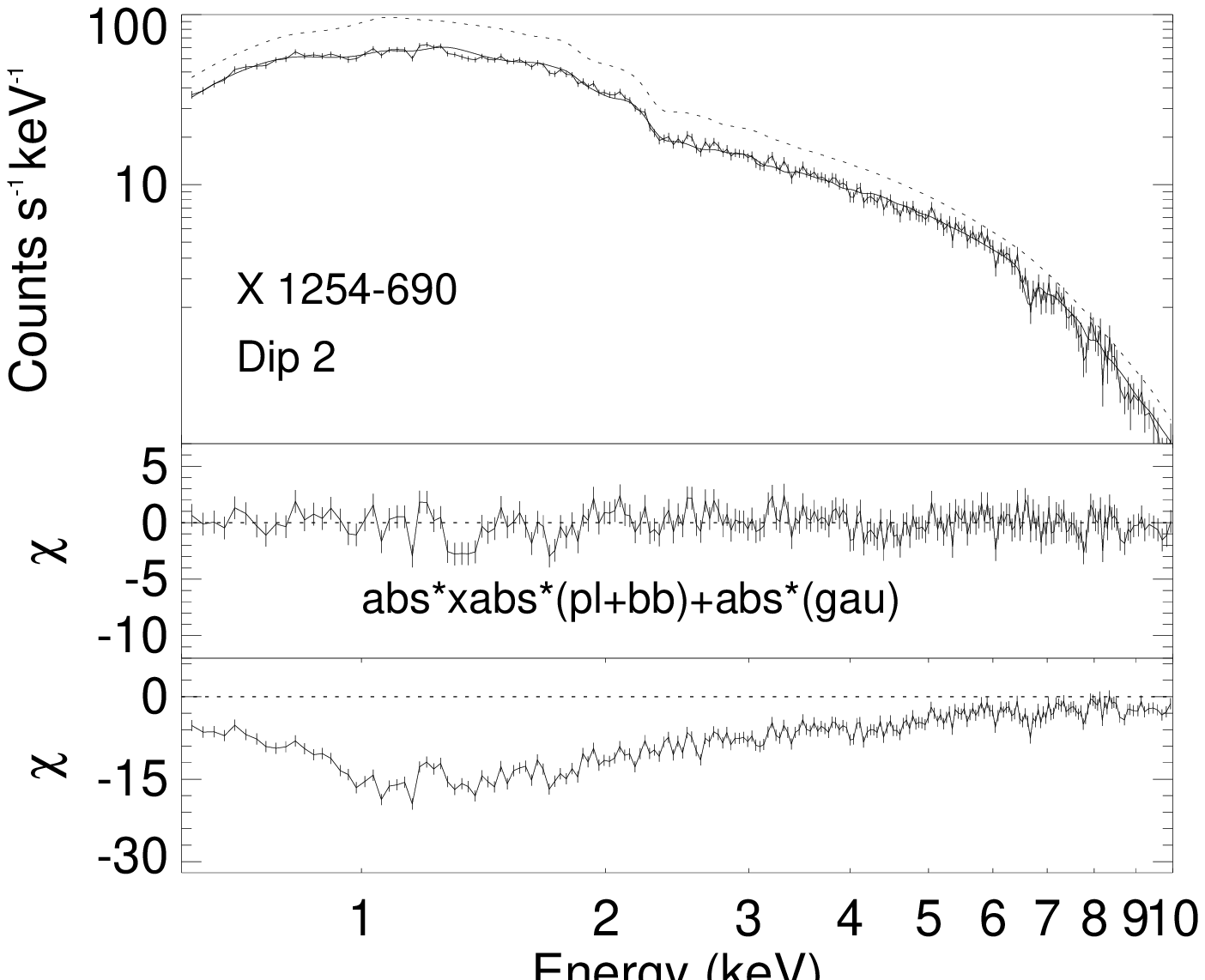}
\hspace{-2.3cm}
\includegraphics[width=0.47\textwidth]{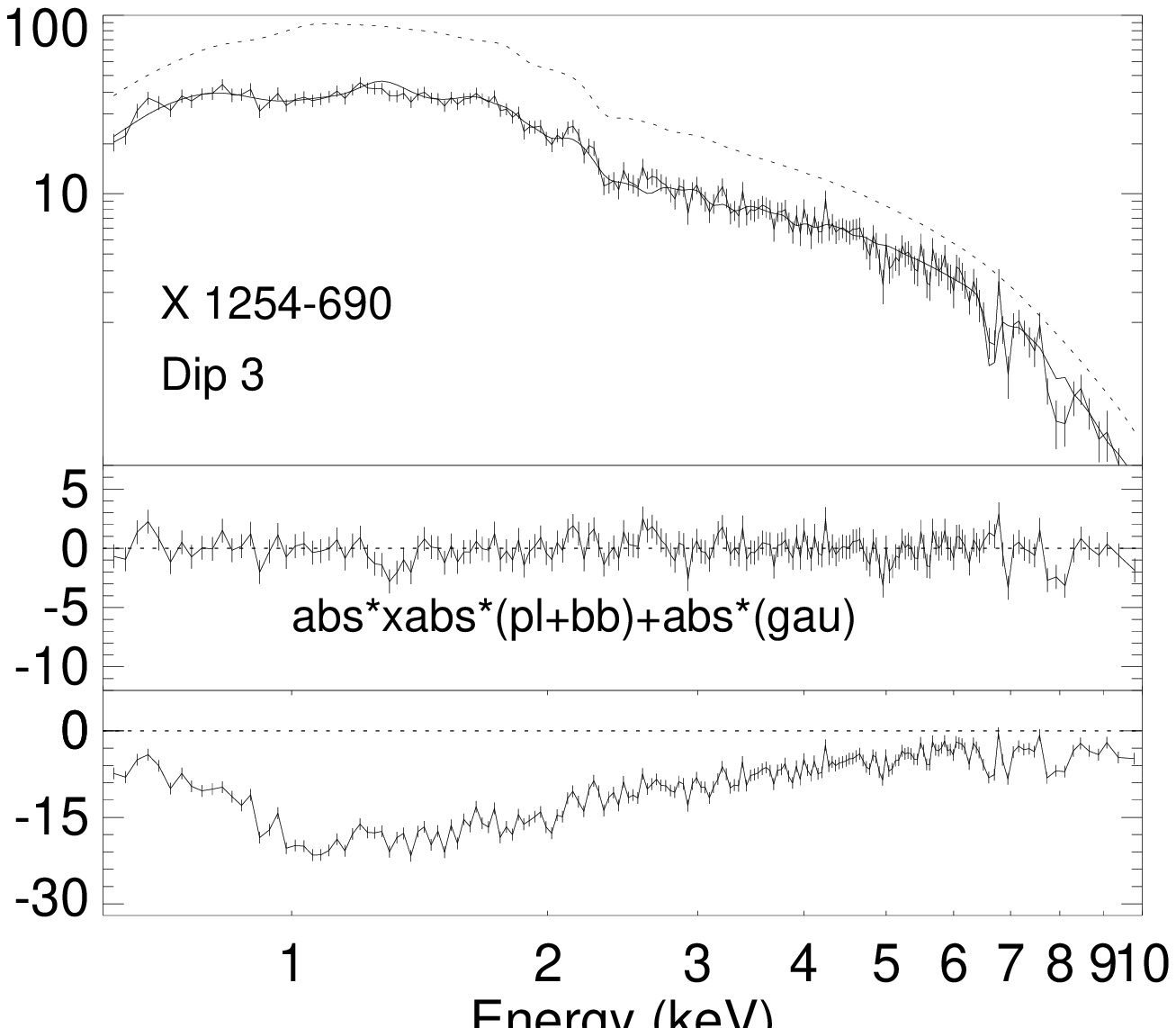}
} \caption{\twelve: EPIC pn persistent and 3 dipping spectra fit
with a power-law ({\tt pl}) and blackbody model
({\tt bb}), modified by absorption from neutral ({\tt abs}) and
ionized ({\tt xabs}) material and 2 narrow emission
lines ({\tt gau}), modified by absorption from neutral material
({\tt abs}) (see Table~\ref{tab:bestfit-1254}).
The dotted lines show the models when the column of the
ionized absorber is set to 0. The middle panels
show the residuals in units of standard deviations from the above
model. The lower panels show residuals when \nhxabs\ is set to 0.
}
\label{fig:bestfit-1254-2}
\end{figure*}

\subsection{\twelve}
\label{sec:twelve}

The XMM-Newton lightcurve of \twelve\ shows evidence for enhanced
background counting rates at the time of occurrence of the dip, with
background peak count rates $\lesssim$10~s$^{-1}$, compared to a
source count rate of 90~s$^{-1}$ during deepest dipping.  For this
reason \citet{1254:boirin03aa} did not examine the spectral changes
during the dip seen from \twelve.  Since this is the only XMM-Newton
observation of a dip from \twelve, we included the dip data in our
analysis. We caution that even though background subtraction is not
critical for such a bright source, the spectral fits may be affected
by the enhanced background.  The best-fits of the persistent and
dipping emission of \twelve\ to the {\tt abs*xabs*(pl+bb) + abs*(gau)}
SPEX model are shown in Fig.~\ref{fig:bestfit-1254-2} and the
parameter values given in Table~\ref{tab:bestfit-1254}. The changes in
the continuum can be modeled primarily by changes in the ionized
absorber with \nhxabs\ increasing from $(8.4 \pm 0.3) \times 10^{22}$
to $(47 \pm 3) \times 10^{22}$~atom~cm$^{-2}$ while \logxi\ decreases
from $4.3 \pm 0.1$ to $2.94 \pm 0.05$ with a smaller increase in the
amount of neutral absorption, from (0.346 $\pm$ 0.002$) \times
10^{22}$ to (0.39 $\pm$ 0.01$) \times 10^{22}$~atom~cm$^{-2}$.

The deepest line in the persistent spectrum is identified with
\fetsix~1s-2p \lya\ with an \ew\ of $32$~eV and is superposed on a
broad emission feature with an energy of 6.8
$\pm$ 0.1~keV and a normalization of (0.08 $\pm$ 0.01$) \times
10^{44}$~photon~s$^{-1}$ (see
Fig.~\ref{fig:bestfit-1254-2}).
The second strongest line
is identified with a mixture of \fetsix~1s-2p \lyb\ and \niteight\
with an \ew\ of $8$~eV.
During dipping, the \ew\ of the \fetsix\ line decreases
and lines of less-ionized species of Fe, such as \fetfive, appear.


\subsection{\fouru}
\label{sec:fouru}

\begin{figure}
\vspace{-0.2cm}
\centerline{\includegraphics[angle=0,width=0.48\textwidth]{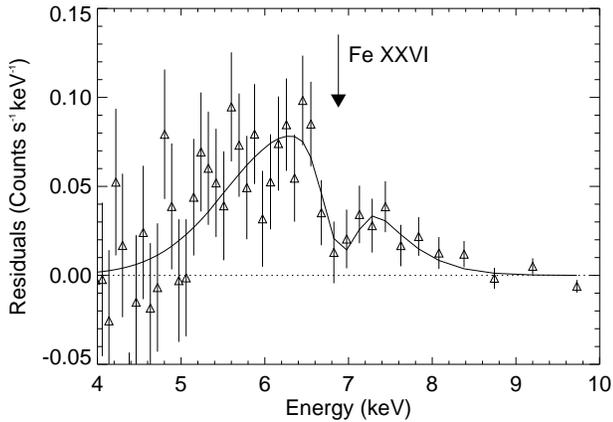}}
\caption{\fouru: 4-10 keV persistent emission spectral residuals from
the best-fit {\tt wabs(bb+pl+gau+gau)} model (see Table
\ref{tab:fouru-xspec}) when the normalizations of the Gaussian
emission feature and the Gaussian absorption feature are set to
zero. The energy of the \fetsix\ absorption line at redshift 0 is
indicated.}
\label{fig:lineres-1746}
\end{figure}

No narrow X-ray absorption features have been reported for this
source.  The observation starts during a dip and two more complete
dips are present in the lightcurve (Fig.~\ref{fig:lightcurves}).  The
dips are very shallow (85\% of the persistent emission remains) and
difficult to identify. We find a dip recurrence period of
$\sim$5.1~hr, in agreement with the value of 5.16~$\pm$~0.01~hr
reported in \citet{1746:balucinska04mn}.  The observation was divided
into persistent and dip spectra which were fit separately using an
absorbed power-law and blackbody {\tt wabs(bb+pl)} model. The fit
quality is acceptable with {\rchisq}s of 1.32 and 1.23 for 229 d.o.f
for persistent and dip spectra, respectively. Examination of the
residuals reveals a broad emission feature at an energy of
$\sim$6.4~keV in both spectra. This feature was modeled as a Gaussian
emission line with energies of $6.45 \, ^{+0.25}_{-0.29}$~keV and $6.1
\, ^{+0.3}_{-0.2}$~keV, widths ($\sigma$) of 0.85 keV and $EW$s of 181
$\pm$ 55~eV and 175 $\pm$ 60~eV for the persistent and dip spectra,
respectively. This results in {\rchisq}s of 1.23 and 1.14 for 226
d.o.f. The F-statistic values of 6.33 and 7.38 indicate that the
probability of such a decrease occurring by chance is $4 \times
10^{-4}$ and $1 \times 10^{-4}$. A Gaussian absorption line with an
energy of $6.9 \pm 0.1$~keV, a width ($\sigma$) of $<$0.80~keV and an
$EW$ of $30 \pm 15$~eV is the best-fit to the absorption feature in
the persistent spectrum. This results in a \rchisq\ of 1.21 for 223
d.o.f. The F-statistic value of 2.19 indicates that the probability of
such a decrease occurring by chance is only
0.09. Table~\ref{tab:fouru-xspec} gives the best-fit parameters and
the residuals are shown in Fig.~\ref{fig:lineres-1746}.

The \fetsix\ narrow absorption feature indicates the presence of a
highly-ionized absorber.  The upper limit $EW$ to a narrow feature at
the energy of \fetfive\ is $<$5~eV. We then fit the persistent and dip
spectra simultaneously with the {\tt abs*xabs*(pl+bb)+abs*(gau+gau)}
model of SPEX. Unfortunately, due to the shallowness of the \fetsix\
feature the highly-ionized absorber is not well constrained and the
best-fit values of \nhxabs\ are highly correlated with variations in
the normalization of the continuum.


\subsection{\mxb}
\label{sec:mxb}

\begin{figure*}[ht!]
\centerline{\includegraphics[angle=0,width=0.47\textwidth]{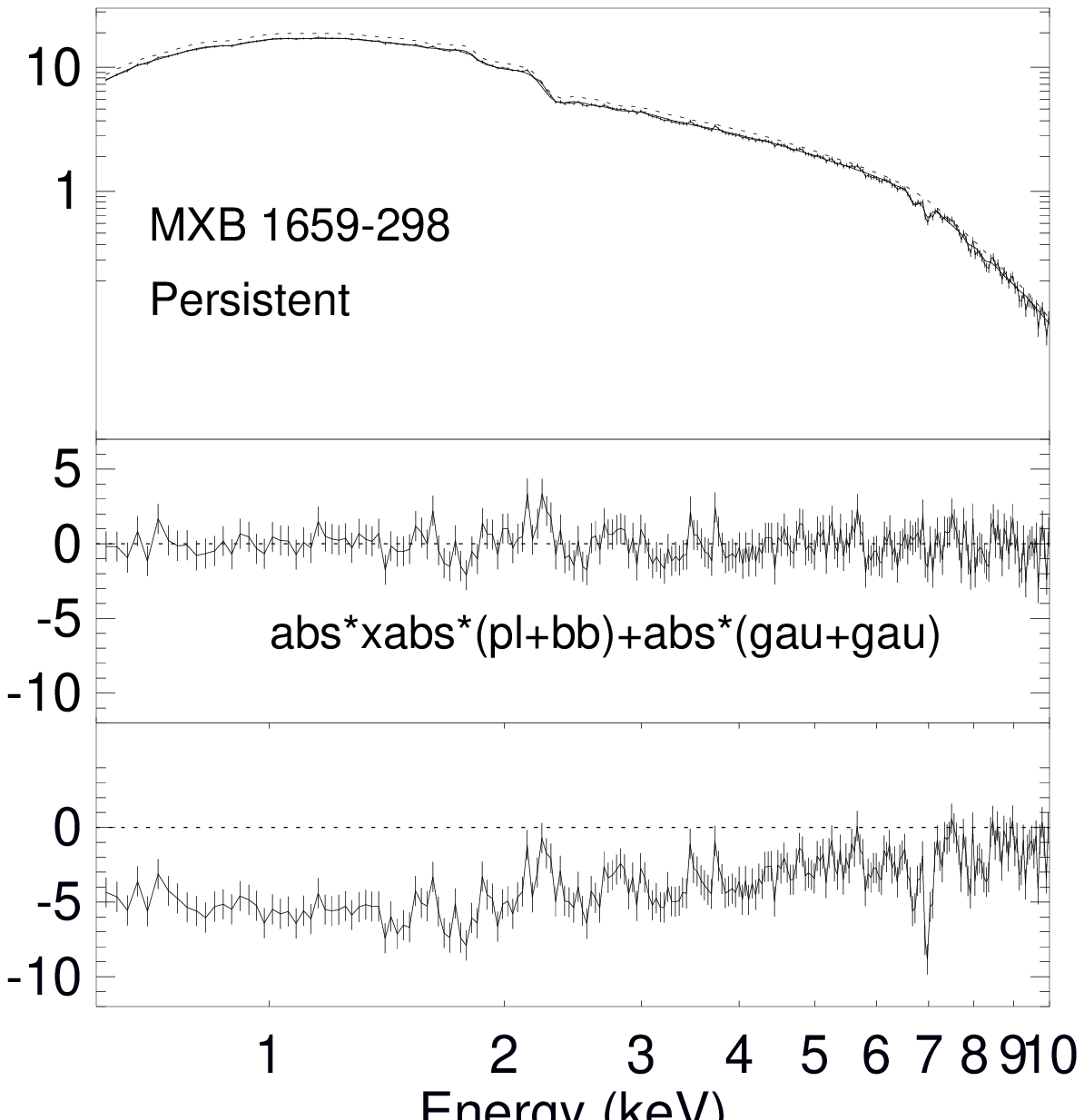}
\hspace{-2.8cm}
\includegraphics[width=0.47\textwidth]{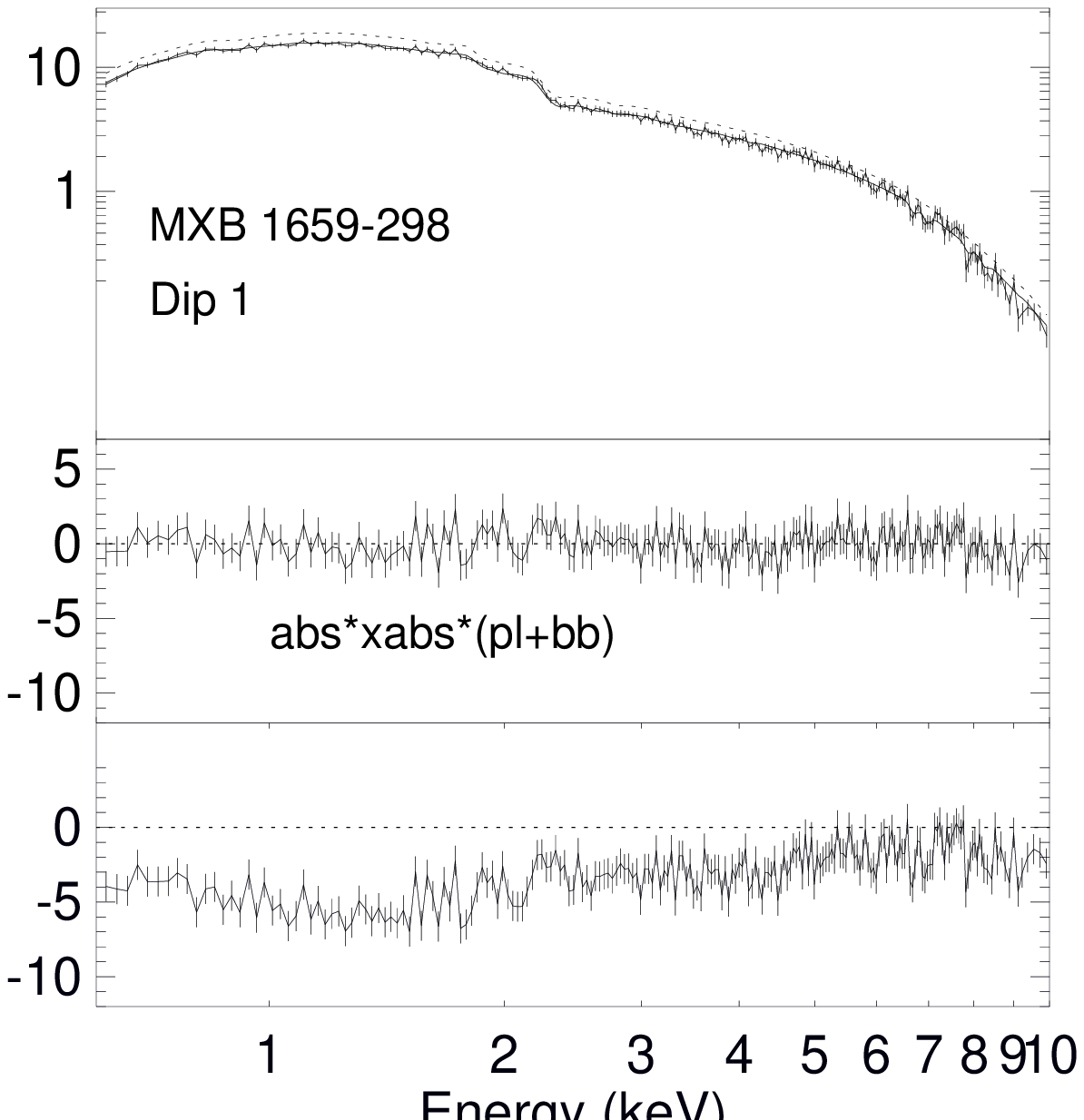}
\hspace{-2.8cm}
\includegraphics[width=0.47\textwidth]{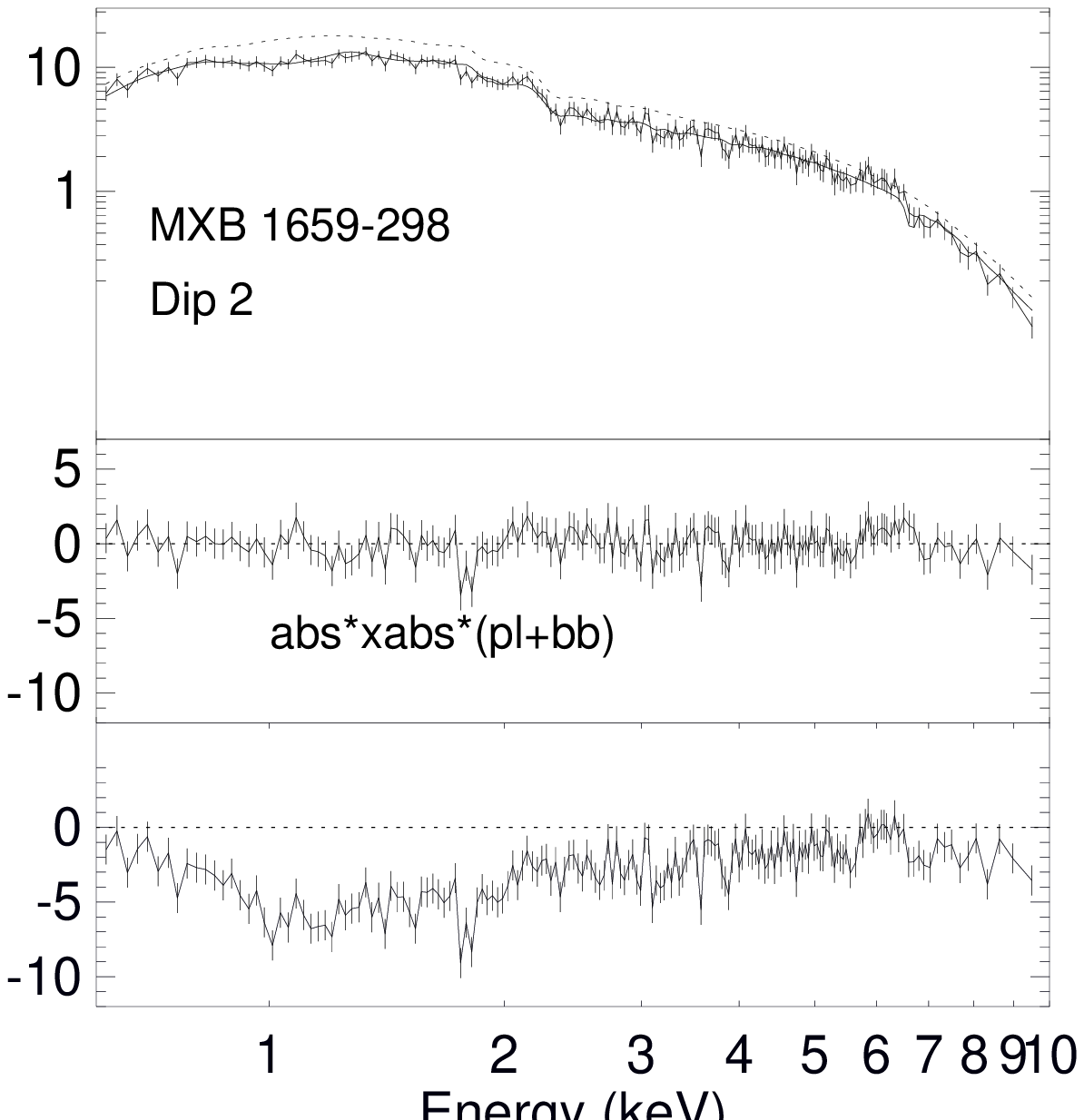}
} \vspace{0.0cm} \centerline{
\includegraphics[width=0.47\textwidth]{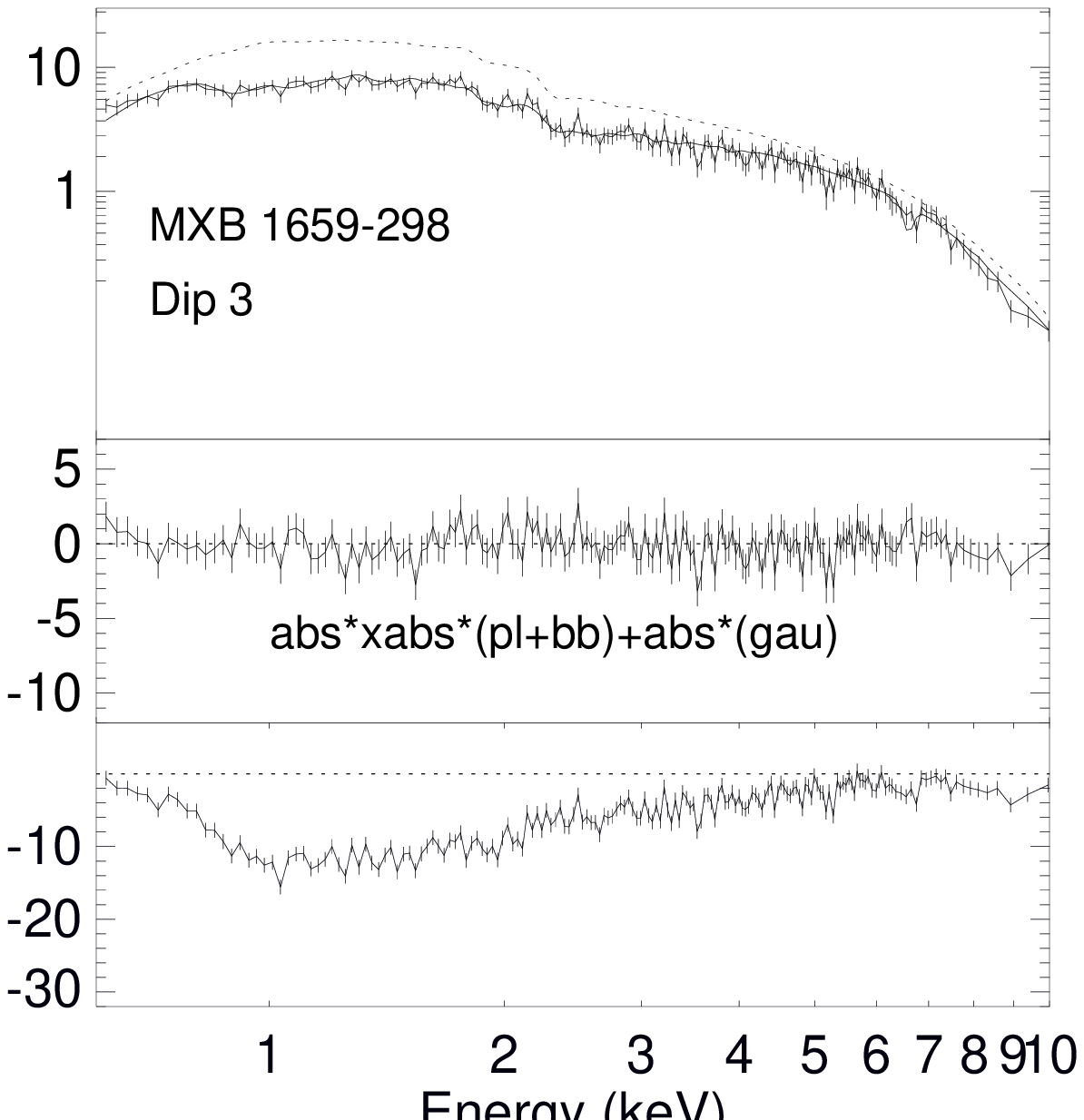}
\hspace{-2.8cm}
\includegraphics[angle=0,width=0.47\textwidth]{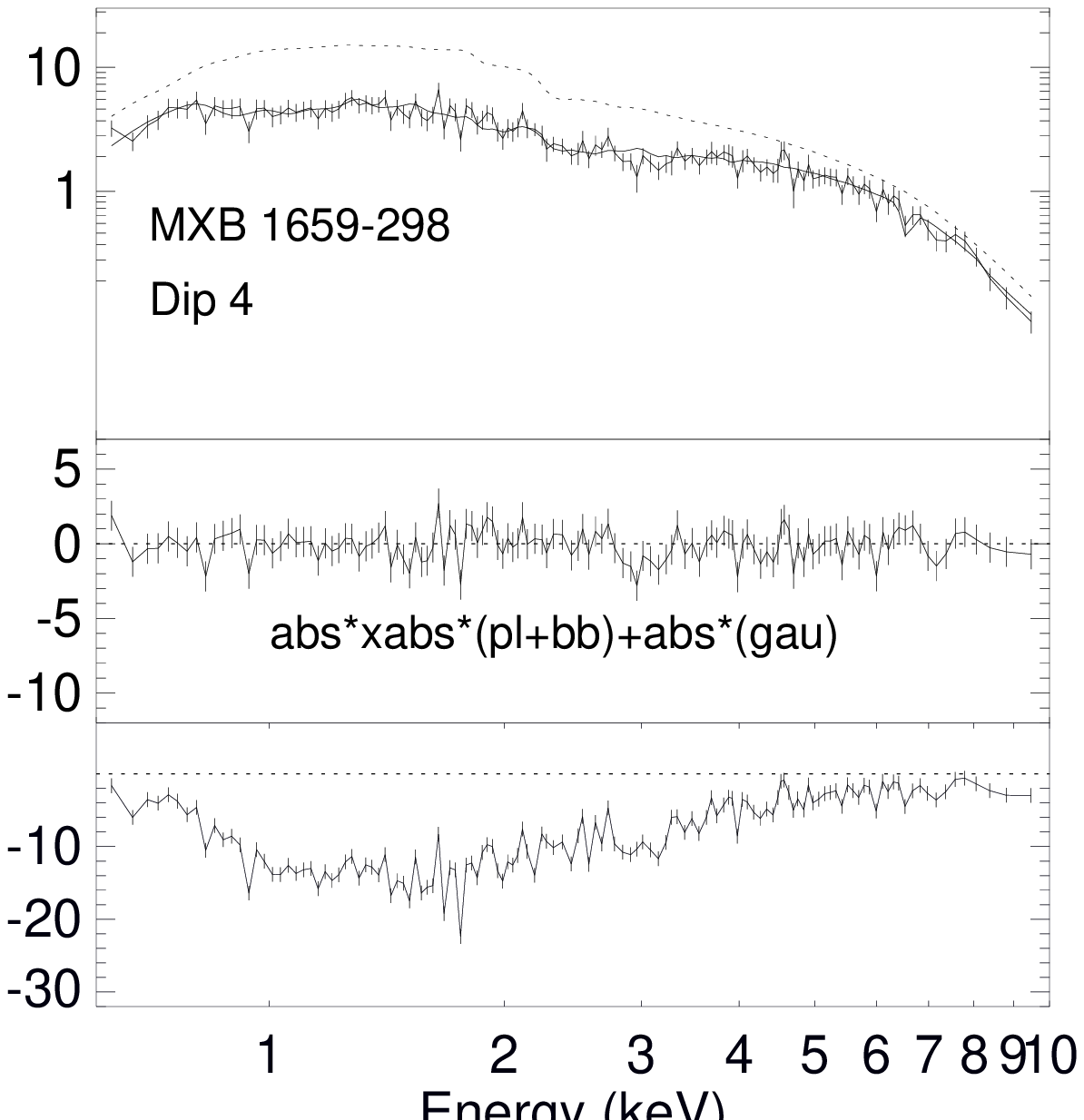}
\hspace{-2.8cm}
\includegraphics[width=0.47\textwidth]{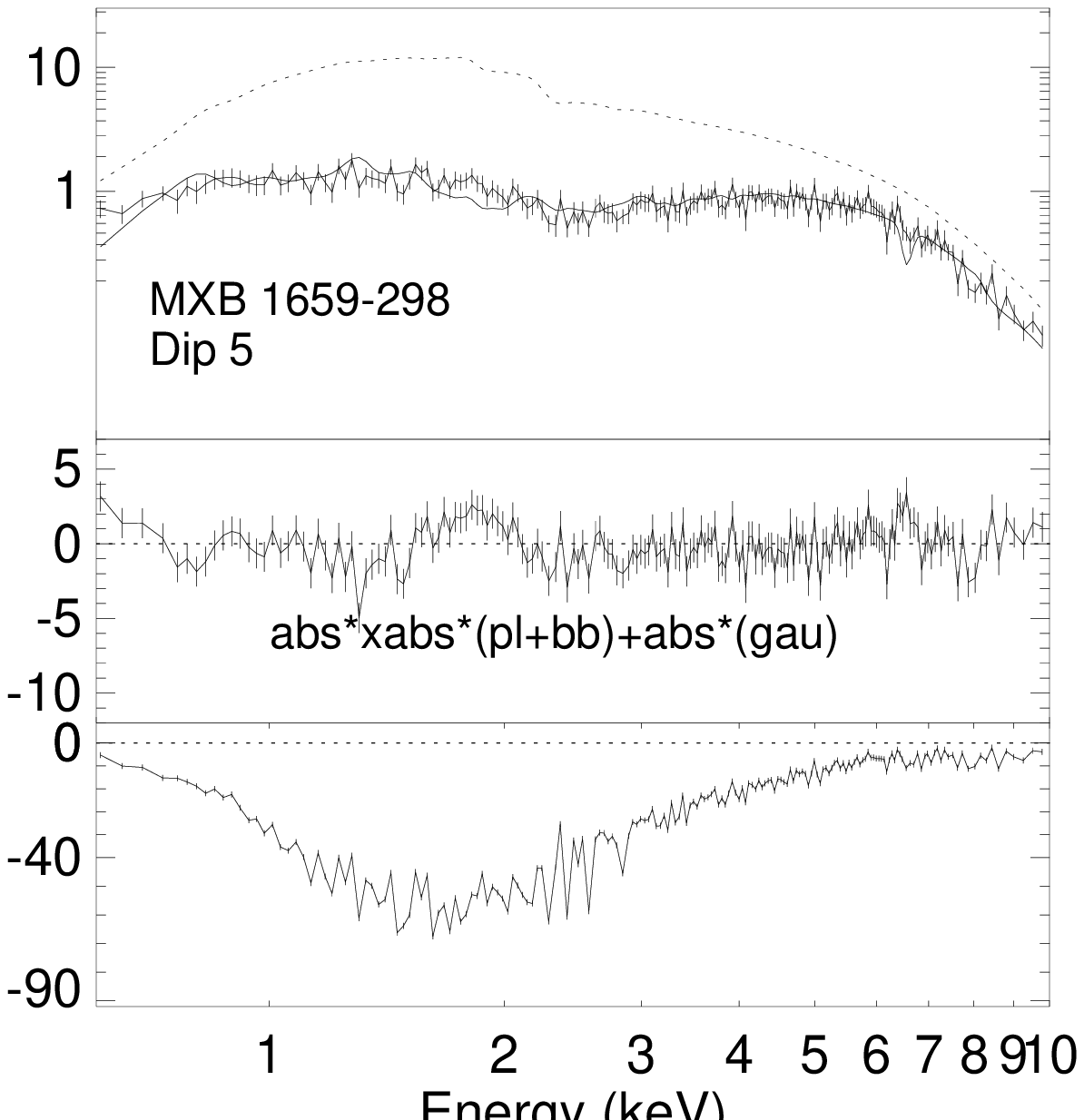}
}
\vspace{0.2cm}
\caption{\mxb: EPIC pn persistent and dipping spectra fit with a
power-law ({\tt pl}) and blackbody model ({\tt bb}), modified by
absorption from neutral ({\tt abs}) and ionized ({\tt xabs}) material
and 2 narrow emission lines ({\tt gau}), modified by absorption from
neutral material ({\tt abs}) (see Table~\ref{tab:bestfit-1659}). The
dotted lines show the models when the column of the ionized absorber
is set to 0. The middle panels show the residuals in units of standard
deviations from the above model.  The lower panels show residuals when
\nhxabs\ is set to 0.}
\label{fig:bestfit-1659-2}
\end{figure*}

The best-fit parameters of the \mxb\ persistent and dipping
emission to the {\tt abs*xabs*(pl+bb) + abs*(gau+gau)} SPEX model are shown in Table~\ref{tab:bestfit-1659} and the
spectra and residuals in Fig.~\ref{fig:bestfit-1659-2}.
The changes in
the continuum can be modeled primarily by a change in the ionized
absorber with \nhxabs\ increasing from (11.1 $\pm$ 0.6$) \times
10^{22}$ to (53 $\pm$ 3$) \times 10^{22}$~atom~cm$^{-2}$ while
\logxi\ decreases from 3.8 $\pm$ 0.1 to $2.42\,^{+0.02}_{-0.06}$ with a
smaller change in the amount of neutral absorption, \nh\ which
increases from (0.306 $\pm$ 0.003$) \times 10^{22}$ to
(0.71 $\pm$ 0.04$) \times 10^{22}$~atom~cm$^{-2}$.

The $\sim$1~keV feature is not visible in the Dip 1 and 2 spectra. The
$\sim$7~keV broad emission feature is only present in the persistent
spectrum.  The strongest absorption lines in the persistent spectrum
are identified with \fetfive\ and \fetsix\ with \ews\ of 32~eV and
48~eV, respectively (see Fig.~\ref{fig:bestfit-1659-2}).  \fetfive\
and \fetsix\ \lyb\ absorption features are also predicted by the {\tt
xabs} model with \ews\ of 9~eV and 12~eV, respectively.  During
dipping, the \ew\ of the \fetsix\ line decreases and the feature
observed at an energy of $\sim 6.60$~keV is identified with a blend of
\fetfive\ to \fenineteen\ lines.

From all the studied sources \mxb\ shows the deepest \fetfive\ and
\fetsix\ absorption lines and is the only dipping LMXB which shows
strong narrow absorption lines in the RGS \citep[][]{1658:sidoli01aa}.
The latter features are produced by \oeight\ 1s-2p, 1s-3p, 1s-4p and
\neten\ 1s-2p, at 18.95$\,^{+0.02}_{-0.01}$,
16.00$\,^{+0.02}_{-0.01}$, 15.21 $\pm$ 0.01 and
12.15$\,^{+0.02}_{-0.01}$ \ang\ and indicate the presence of less
ionized material than that responsible for the \fetfive\ and \fetsix\
absorption features. Thus, the absorbing material may have a range of
ionization states: the material closer to the compact object may be
more ionized and produce the \fetfive\ and \fetsix\ lines while the
one farther from the compact object may be less ionized and produce
the \oeight\ and \neten\ lines.  The latter features are however not
predicted by the {\tt xabs} model. This is probably because when
fitting an absorber with only one ionization state, the \fetfive\ and
\fetsix\ lines dominate over the \oeight\ and \neten\ lines due to
their larger $EW$s. We next performed a fit to the persistent RGS
first order spectra with the SPEX {\tt abs*xabs*(pl)} model. The RGS
dipping spectra have too few counts for spectral fitting. We find \nh\
= (0.287 $\pm$ 0.007$) \times 10^{22}$~atom~cm$^{-2}$, \nhxabs\ = $(3
\, ^{+5}_{-2}) \times 10^{22}$~atom~cm$^{-2}$, \logxi\ = 3.1 $\pm$ 0.3
and $\sigma_v$ = 190 $\, ^{+180}_{-100}$~km~s$^{-1}$. The
uncertainties in these parameters are large in comparison with those
found for the pn fit.  Despite the uncertainties, a less ionized
absorber is required by the RGS data compared to the pn.  Narrow
absorption features of \oeight\ 1s-2p, 1s-3p, 1s-4p and \neten\ 1s-2p
are predicted by the {\tt xabs} model with \ews\ of 2.2, 0.7, 0.3 and
2.0~eV, respectively. These values are slightly lower than those found
by \citet[][]{1658:sidoli01aa} of 2.6 $\pm$ 0.4, 1.3 $\pm$ 0.4, 1.6
$\pm$ 0.4 and 2.1 $\pm$ 0.4~eV for the corresponding lines. This
difference may be explained by our use of a complete photo-ionized
absorber rather than individual spectral features. Thus, the
assumption of a range of different ionization states for the absorber
seems correct. However, when we added a second absorber to our model
with \logxi\ fixed to the RGS value and performed a fit to the pn
persistent spectrum, the \chisq\ of the fit did not improve
significantly. This suggests that the nature of the absorber is even
more complex. Alternatively, the \oeight\ and \neten\ features may be
too weak in the pn spectrum to constrain properly the second ionized
absorber, since such features have been reported only in the RGS data,
which has a resolution ($E/\Delta E)$ of $\sim$300 compared to
$\sim$12 for the pn at 1~keV.
\begin{table}
\caption{Best-fit spectral parameters for the persistent \fouru\
emission with the {\tt wabs(bb+pl+gau+gau)} model. The $\sigma$ of the
$\sim$6.4~keV Gaussian emission line is constrained to be
$\le$0.85~keV.  }\begin{tabular}{l@{\extracolsep{0.15cm}}l@{\extracolsep{0.15cm}}c@{\extracolsep{0.15cm}}}
\hline \hline\noalign{\smallskip}
Comp. & Parameter & Value \\
\hline\noalign{\smallskip}
{\tt wabs}& \nh  {\small($10^{22}$ cm$^{-2}$)} & 0.37 $\pm$ 0.01 \\
{\tt pl} & \phind\ & 1.69 $\pm$ 0.03 \\
& \kpl\ {\small(photon keV$^{-1}$~cm$^{-2}$~s$^{-1}$)} & 0.115 $\pm$ 0.003 \\
{\tt bb} & $kT_{bb}$ (keV) & 1.28 $\pm$ 0.04 \\
& \kbb\ {\small($L_{39}/d_{10}^2$)} & (3.6 $\pm$ 0.2) $\times 10^{-3}$\\
Emission & \egau\ {\small(keV)}               & 6.45$\,^{+0.25}_{-0.29}$ \\
 & $\sigma$ {\small(keV)}               &  0.85       \\
& \ew\ (eV) &  181 $\pm$ 55 \\
\fetsix\ & \egau\ {\small(keV)}  &    6.9 $\pm$ 0.1          \\
absorption& $\sigma$ {\small(keV)}               & $<$0.80        \\
& \ew\ (eV) &  30 $\pm$ 15  \\
\noalign{\smallskip\hrule\smallskip}
\label{tab:fouru-xspec}
\end{tabular}
\end{table}

\begin{table*}
\caption{\mxb: best-fits to the EPIC pn persistent and 5 dip
spectra using the {\tt abs*xabs*(pl+bb)+abs*(gau+gau)} model. The
components of the continuum {\tt (pl+bb)} are the same for all
spectra and only the \nh\ of the neutral absorber, the line
parameters, \nhxabs, $\sigma _v$ and $\xi$ of the ionized absorber
are individually fit for each spectrum. \flux\ is the 0.6--10 keV
absorbed flux.  The $FWHM$ of the 6.4~keV Gaussian emission line is
constrained to be $\le$2.0~keV. The $FWHM$s of the $\sim$1~keV Gaussian
emission lines are constrained to be $\le$0.2~keV.}
\begin{center}
\begin{tabular}{lccccccc}
\hline \hline\noalign{\smallskip}
 & & Persistent & Dip 1 & Dip 2 & Dip 3 & Dip 4 & Dip 5 \\
\noalign{\smallskip\hrule\smallskip}
& Comp. & & & & & &  \\
Parameter & & & & & & &  \\
& {\tt pl} & & & &  \\
\phind & & \multicolumn{5}{c}{1.96 $\pm$ 0.03}  \\
\multicolumn{2}{l}{\kpl\ {\small ($10^{44}$ ph. s$^{-1}$ keV$^{-1}$)}} & \multicolumn{5}{c}{51 $\pm$ 2} \\
& {\tt bb} & & & &\\
\ktbb\ {\small(keV)} & & \multicolumn{5}{c}{1.14 $\pm$ 0.03}  \\
\kbb\  {\small($10^{11}$ cm$^{2}$)} & & \multicolumn{5}{c}{59 $\pm$ 5}  \\
& {\tt abs} & & & & & & \\
\multicolumn{2}{l}{\nh\ {\small($10^{22}$ cm$^{-2}$)}} & 0.306 $\pm$ 0.003 & 0.302 $\pm$ 0.005 & 0.34 $\pm$ 0.01 & 0.40 $\pm$ 0.02 & 0.47 $\pm$ 0.04 & 0.71 $\pm$ 0.04  \\
& {\tt xabs} & & & & & &  \\
\multicolumn{2}{l}{\nhxabs\ {\small($10^{22}$ cm$^{-2}$)}} & 11.1 $\pm$ 0.6 & 20 $\pm$ 2 & 19 $\pm$ 4 & 18 $\pm$ 3 & 24 $\pm$ 3 & 53 $\pm$ 3 \\
\logxi\ {\small(\xiunit)} & & 3.8 $\pm$ 0.1 & 3.7 $\pm$ 0.2 & 2.9 $\pm$ 0.1 & 2.55$\,^{+0.08}_{-0.06}$ & 2.42$\,^{+0.06}_{-0.09}$ & 2.42$\,^{+0.02}_{-0.06}$  \\
\sigmav\ {\small(km s$^{-1}$)} & & 700$\,^{+1000}_{-350}$ & 150$\,^{+170}_{-100}$ & 160$\,^{+100}_{-70}$ & 140$\,^{+90}_{-40}$ & 140$\,^{+100}_{-70}$ & 110$\,^{+40}_{-30}$   \\
\noalign {\smallskip}
& {\tt gau}  & & & & & & \\
\multicolumn{2}{l}{ \egau\ {\small(keV)}} & 6.4 $\pm$ 0.2 & & & & & \\
\multicolumn{2}{l}{ $FWHM$ {\small(keV)}} & 2 &  & & & & \\
\multicolumn{2}{l}{ \kgau\ {\small(10$^{44}$ ph s$^{-1}$)}} & 0.53 $\pm$ 0.08 & & & & &  \\
\noalign {\smallskip}
& {\tt gau}  & & & & & & \\
\multicolumn{2}{l}{ \egau\ {\small(keV)}}               & 0.99 $\pm$ 0.03  & & & 0.96 $\pm$ 0.03 & 0.92 $\pm$ 0.03 & 0.93 $\pm$ 0.02     \\
\multicolumn{2}{l}{ $FWHM$ {\small(keV)}}               & 0.11$\,^{+0.09}_{-0.07}$ & & & 0.15 $\pm$ 0.09& 0.2 & 0.2 \\
\multicolumn{2}{l}{ \kgau\ {\small(10$^{44}$ ph s$^{-1}$)}} & 0.4 $\pm$ 0.2 & & & 1.7$\,^{+1.0}_{-0.7}$ & 3 $\pm$ 2 & 2.2 $\pm$ 0.6  \\
\noalign {\smallskip}
\hline\noalign {\smallskip}
\multicolumn{2}{l}{\flux\ \small (10$^{-10}$ \ergcms)}  & 9.7 & 8.9 & 7.9 & 6.6 & 5.4 & 2.9  \\
    \multicolumn{2}{l}{\rchisq (d.o.f.)} & \multicolumn{6}{c}{1.22 (1096)} \\
        \multicolumn{2}{l}{Exposure (ks)} & 11.8 & 2.16 & 0.48 & 0.56 & 0.35 & 1.33  \\
\noalign{\smallskip\hrule\smallskip}
\label{tab:bestfit-1659}
\end{tabular}
\end{center}

\end{table*}

\subsection{\bigdip}
\label{sec:bigdip}

\begin{table*}
\caption{\bigdip: best-fits to the EPIC pn 5 dip spectra using the
{\tt abs$_{1}$*xabs*e$^{-\tau}$(bb+pl) +
abs$_{2}$*(1-e$^{-\tau}$)*(bb+pl)} model. The components of the
continuum {\tt (pl+bb)}, the column density of the halo neutral
absorber and $\tau$ are the same for all spectra and only the \nh\
of the neutral absorber, and \nhxabs, $\sigma _v$ and $\xi$ of the ionized
absorber are individually fit for each spectrum. \flux\ is the 0.6--10
keV absorbed flux. \phind\ is constrained to be $\ge$1.90. }
\begin{center}
\begin{tabular}{lcccccc}
\hline \hline\noalign{\smallskip}
 & & Dip 1 & Dip 2 & Dip 3 & Dip 4 & Dip 5 \\
\noalign{\smallskip\hrule\smallskip}
& Comp. & & & & &  \\
Parameter & & & & & & \\
& {\tt pl} & & & &  \\
\phind & & \multicolumn{5}{c}{1.90}  \\
\multicolumn{2}{l}{\kpl\ {\small ($10^{44}$ ph. s$^{-1}$ keV$^{-1}$)}} & \multicolumn{5}{c}{120$\,^{+60}_{-20}$} \\
& {\tt bb} & & & & &\\
\ktbb\ {\small(keV)} & & \multicolumn{5}{c}{1.20$\,^{+0.14}_{-0.09}$}  \\
\kbb\  {\small($10^{11}$ cm$^{2}$)} & & \multicolumn{5}{c}{64$\,^{+30}_{-20}$}  \\
& {\tt abs$_1$} & & & & & \\
\multicolumn{2}{l}{\nh\ {\small($10^{22}$ cm$^{-2}$)}} & 10.7 $\pm$ 0.5 & 11.8 $\pm$ 0.9 & 14 $\pm$ 2 & 31 $\pm$ 2 & 59 $\,^{+6}_{-3}$ \\
& {\tt xabs} & & & & &  \\
\multicolumn{2}{l}{\nhxabs\ {\small($10^{22}$ cm$^{-2}$)}} & 13 $\pm$ 2 & 15 $\pm$ 4 & 27 $\pm$ 4 & 29 $\pm$ 4 & 68 $\pm$ 9 \\
\logxi\ {\small(\xiunit)} & & 3.6 $\pm$ 0.2 & 3.2 $\pm$ 0.3 & 2.9 $\pm$ 0.2 & 3.0 $\pm$ 0.2 & $\ge$3.3 \\
\sigmav\ {\small(km s$^{-1}$)} & & 280$\,^{+250}_{-150}$ & 140$\,^{+200}_{-100}$ & $<$128 & $<$94 & 110$\,^{+200}_{-70}$ \\
\noalign {\smallskip}
& {\tt Halo} & & & & &\\
& {\tt abs$_2$} & & & & & \\
\multicolumn{2}{l}{\nh$^{Halo}$ {\small($10^{22}$ cm$^{-2}$)}} & \multicolumn{5}{c}{10.7 $\pm$ 0.5}  \\
& {\tt etau} & & & & & \\
\multicolumn{2}{l}{$\tau_1$} & \multicolumn{5}{c}{1.8 $\pm$ 0.2}  \\
\noalign {\smallskip}
\hline\noalign {\smallskip}
\multicolumn{2}{l}{\flux\ \small (10$^{-10}$ \ergcms)} &  8.3 & 7.5 & 5.8 & 4.0 & 2.0  \\
        \multicolumn{2}{l}{\rchisq (d.o.f.)} & \multicolumn{5}{c}{1.15 (729)}  \\
        \multicolumn{2}{l}{Exposure (ks)} & 0.3 & 0.1 & 0.2 & 0.4 & 0.9 \\
\noalign{\smallskip\hrule\smallskip}
\label{tab:bestfit-1624}
\end{tabular}
\end{center}

\label{tab:bigdip-bestfit}
\end{table*}

\begin{figure*}[ht!]
\centerline{\includegraphics[angle=0,width=0.47\textwidth]{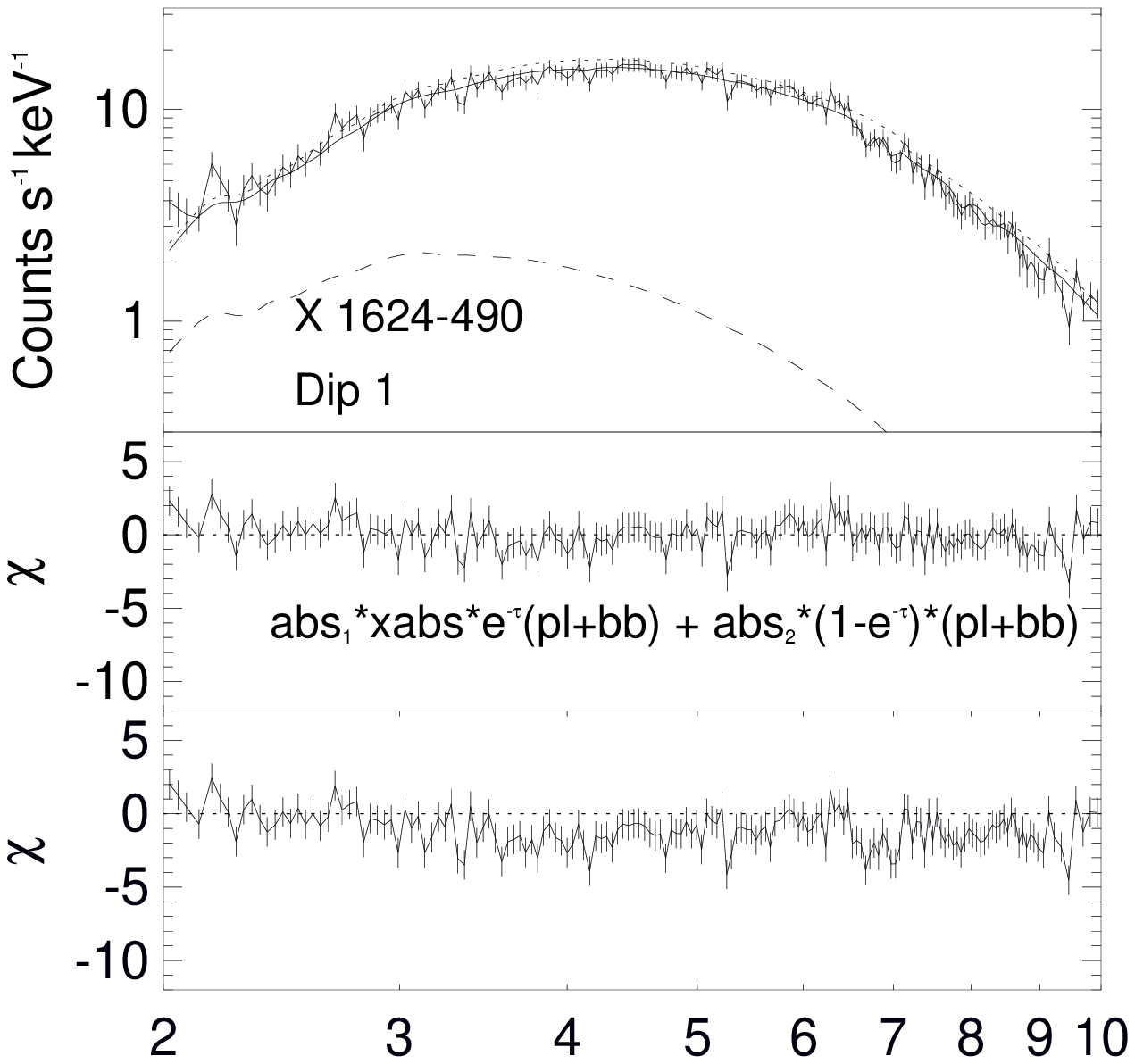}
\hspace{-2.8cm}
\includegraphics[width=0.47\textwidth]{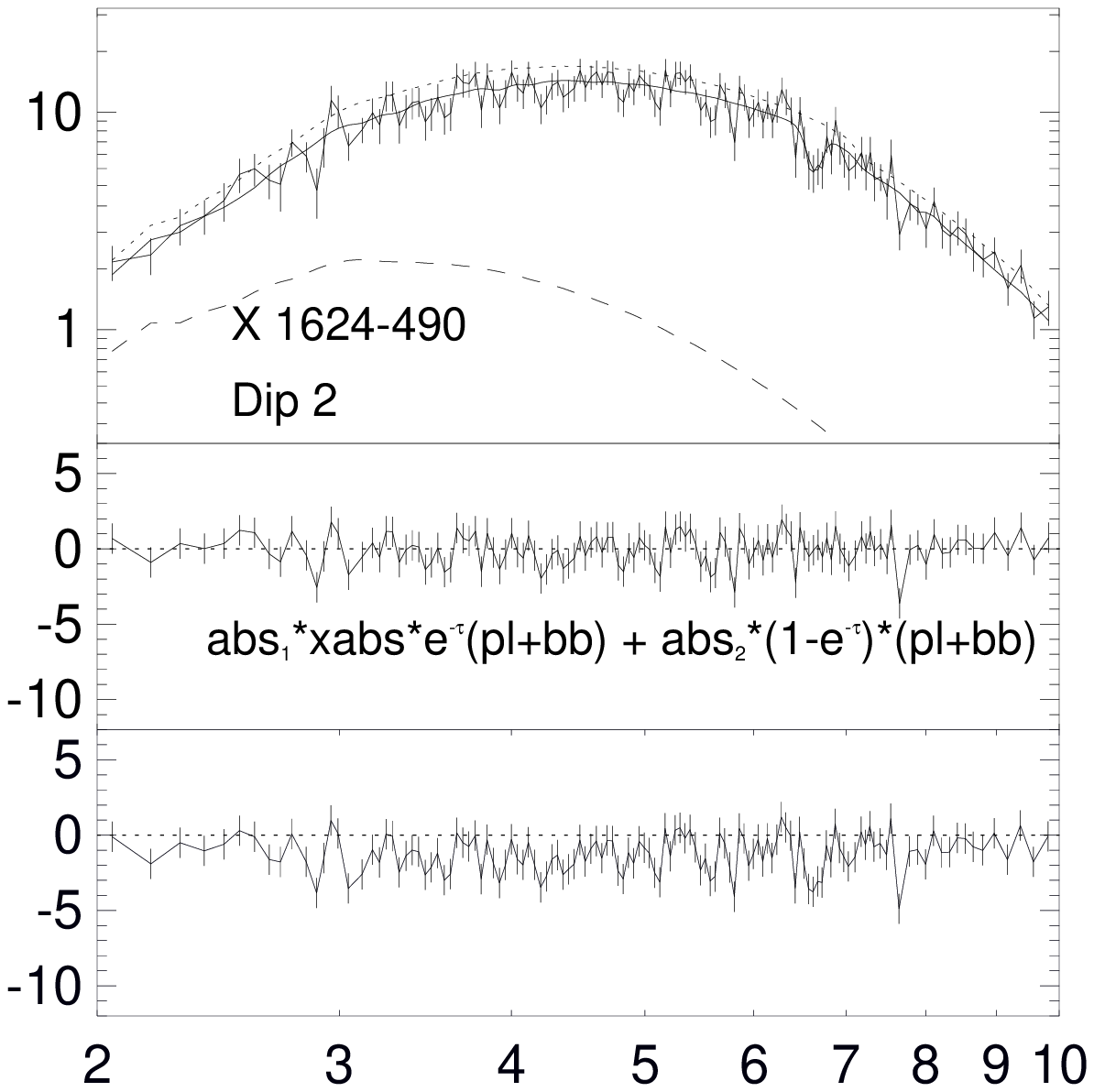}
\hspace{-2.8cm}
\includegraphics[width=0.47\textwidth]{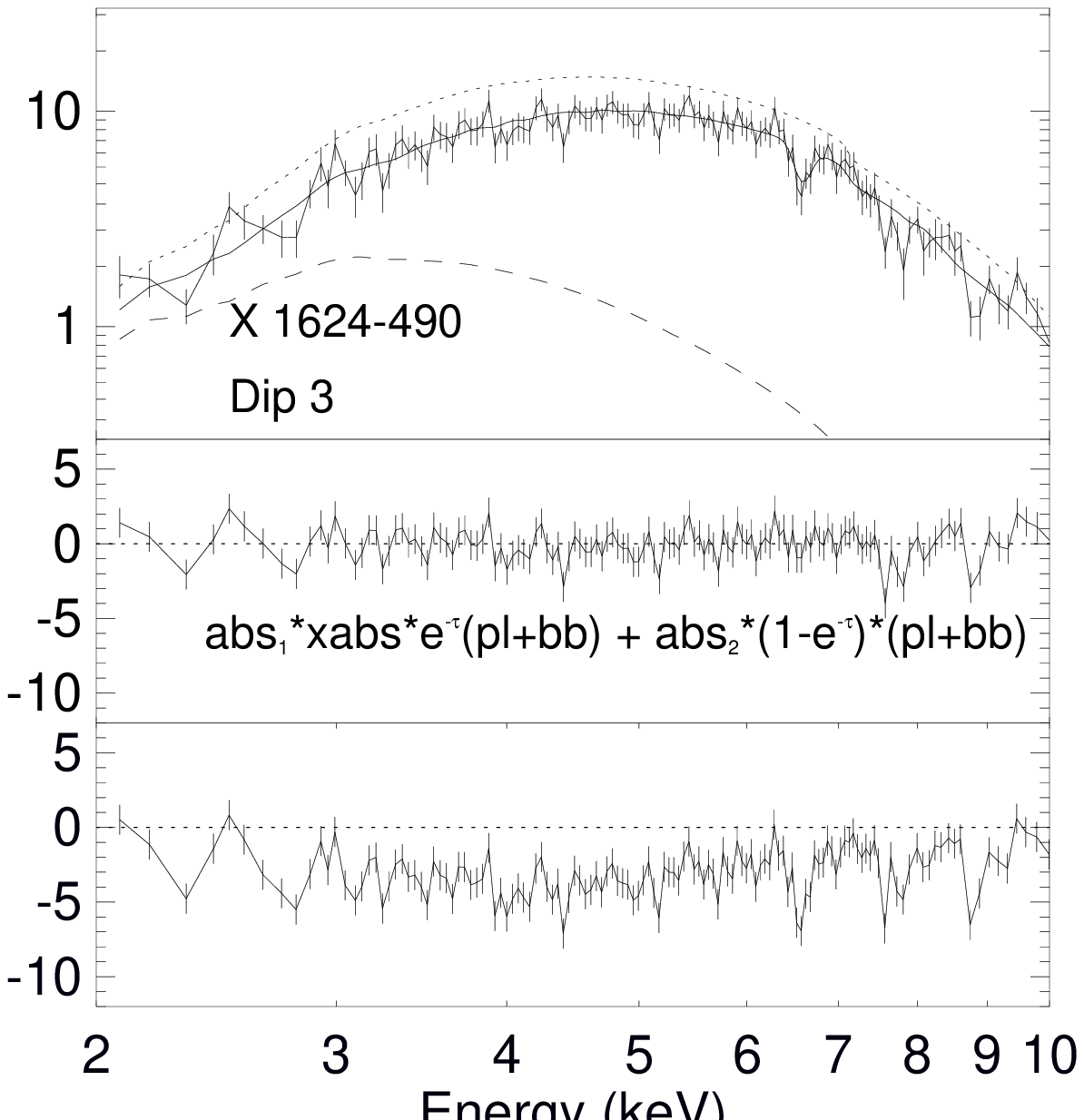}
 \vspace{0.0cm}}
\hspace{-1.2cm}
\includegraphics[angle=0,width=0.47\textwidth]{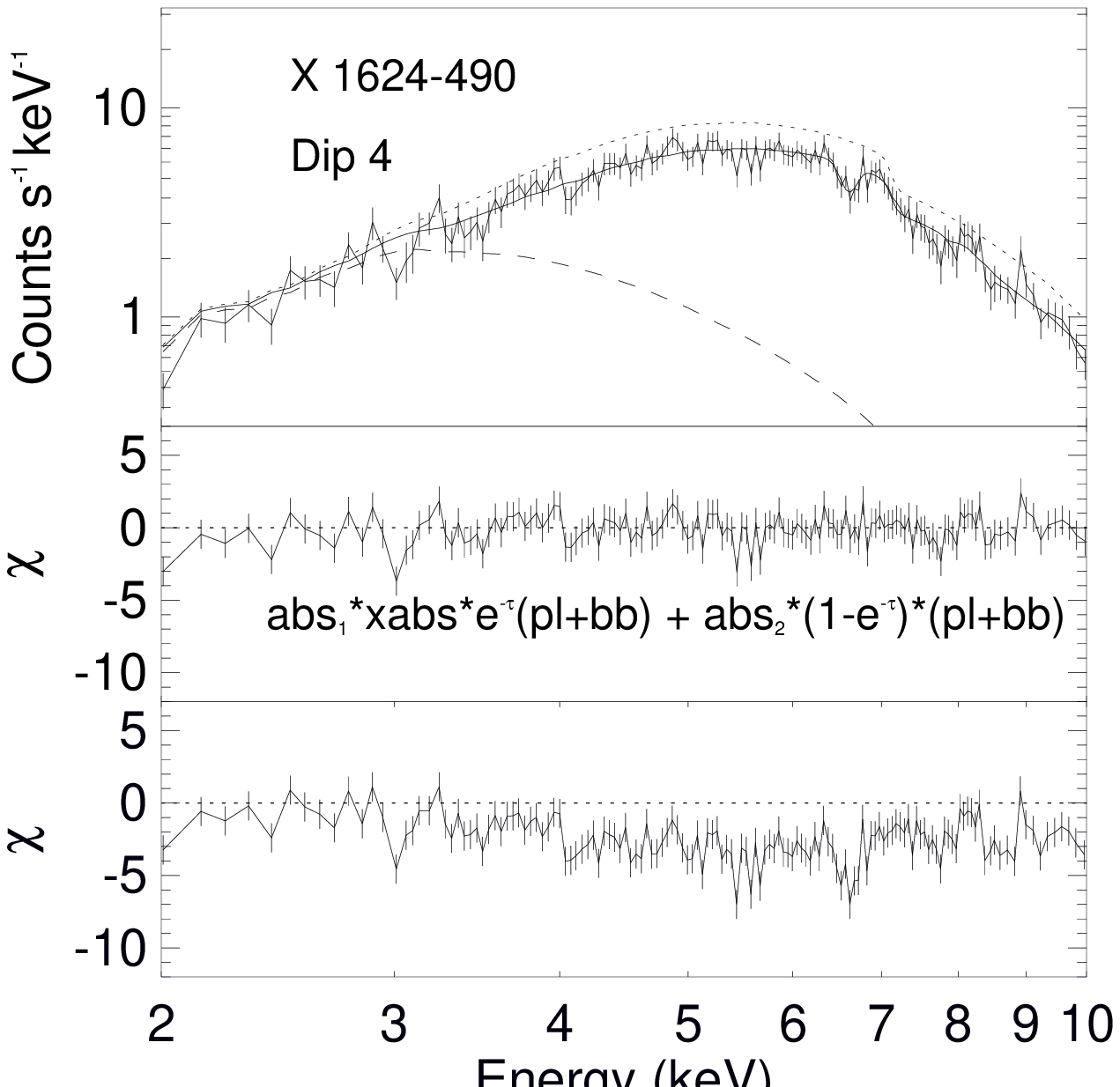}
\hspace{-2.8cm}
\includegraphics[width=0.47\textwidth]{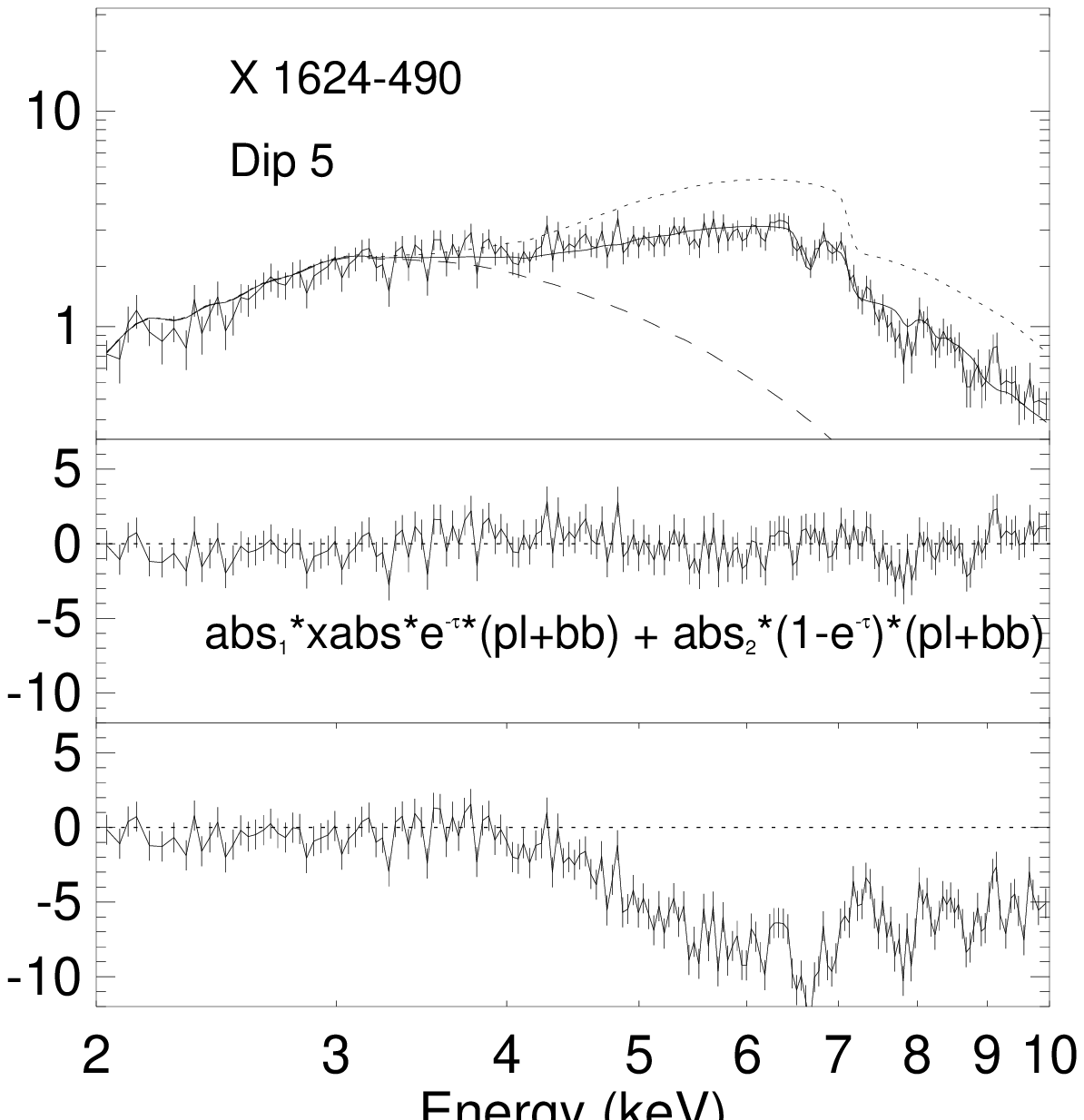}
\vspace{0.2cm}
\caption{\bigdip: EPIC pn dipping spectra fit with the model of
Table~\ref{tab:bestfit-1624}. The dotted lines show the models when
the column of the ionized absorber is set to 0. The dashed lines show
the constant halo component. The middle panels show residuals in units
of standard deviations from the above model. The lower panels show
residuals when the \nhxabs\ of the ionized absorber is set to 0.  }
\label{fig:bestfit-1624-2}
\end{figure*}

The radial intensity profile of the source obtained using the EPIC pn
image revealed a constant excess in intensity at large radii
($>$30\arcsec) during both persistent and dipping intervals,
demonstrating the presence of a dust-scattering halo, first detected
from this source by \citet{1624:angelini97asp}. The galactic column
density of \bigdip\ is high \citep[$\sim$$8 \times
10^{22}$~atom~cm$^{-2}$,][]{1624:parmar02aa}, so that a dust scattered
halo is expected. To model this we used {\tt
abs$_{1}$*xabs*e$^{-\tau}$(bb+pl) +
abs$_{2}$*(1-e$^{-\tau}$)*(bb+pl)}. The first component represents the
contribution of the source reduced by a factor e$^{-\tau}$ 
(component {\tt etau} in SPEX), corresponding to scattering out of
the line of sight.  The scattering into the line of sight due to the
halo is represented by the second component and is expected to be the
same for persistent and dipping spectra. $\tau$ is defined as
$\tau_1$E$^{-2}$, where $\tau_1$ is the optical depth at 1~keV and
E$^{-2}$ is the theoretical energy dependence of the dust scattering
cross-section. Unfortunately, the \bigdip\ persistent spectrum is
strongly piled-up and counts from the inner circle of 20\arcsec\
radius had to be excluded. Thus, the resultant spectra contain a
larger fraction of the halo component. For this reason we decided to
use only the dipping spectra for the fit, since then the inner
20\arcsec\ radius circle does not have to be excluded and there is a
larger contribution from direct source emission.  We imposed a lower
limit of 1.90 to the power-law photon index, \phind, to prevent it
becoming negative without significantly changing the $\chi^2$ of the
fit. The value of 1.90 was taken from \citet{1624:parmar02aa} (their
value of \phind\ = 2.02 $\pm$ 0.12 gives a minimum value of 1.90).
The \nh\ of the neutral absorber for the Dip 1 spectrum was coupled to
the value of \nh\ for the halo, assuming that the source must be at
least as absorbed as the halo.  The best-fits parameters for the
dipping emission of \bigdip\ are shown in Table~\ref{tab:bestfit-1624}
and the spectra and residuals in
Fig.~\ref{fig:bestfit-1624-2}. Significant changes in {\it both} the
ionized and neutral absorbers are required to obtain satisfactory fits
with \nh\ increasing from (10.7 $\pm$ 0.5$) \times 10^{22}$ to $(59 \,
^{+6}_{-3}) \times 10^{22}$~atom~cm$^{-2}$ and \nhxabs\ from (13 $\pm$
2$) \times 10^{22}$ to (68 $\pm$ 9$) \times
10^{22}$~atom~cm$^{-2}$. The value of \logxi\ decreases from 3.6 $\pm$
0.2 to 3.0 $\pm$ 0.2 between Dip 1 and Dip 4. For Dip 5 the value of
\logxi\ is not well constrained and we can only give a lower limit of
$\ge 3.3$. Below $\sim$5~keV the contribution of the halo (dashed
line) dominates during the deepest dip stages. The optical depth of
the dust, $\tau_1$ = 1.8 $\pm$ 0.2, is consistent with that found by
\citet{1624:balucinska00aa} of 2.4 $\pm$ 0.4.

\section{Ionized absorber properties}
\label{sec:phenomenology}

Examination of the spectral fit results (Tables~\ref{tab:1916-bestfit}
to~\ref{tab:bigdip-bestfit}) shows that we are able to successfully
account for the complex changes in the 0.6--10~keV continuum and
absorption lines during dips from the LMXB studied here (with the
exception of \fouru\ where the dips are very shallow) by large
increases in the column density, \nhxabs, and decreases in the amount
of ionization, \xil, of a highly-ionized absorber, together with much
smaller increases in the \nh\ of a neutral absorber (for \bigdip\ the
increase in the column densities of the neutral and ionized absorbers
are comparable). The $EW$s of the lines predicted by \xabs\ are
consistent with previous measurements for the persistent emission (see
Table~\ref{tab:lmxb-lines}).

During the persistent intervals, the values of \sigmav\ are poorly
constrained, except for \exo\ (see Tables~\ref{tab:1916-bestfit}
to~\ref{tab:bigdip-bestfit}). This is evidence for the absorption
lines being unsaturated, i.e. on the linear part of the curve of
growth, where the \ew\ increases linearly with the column density,
independently of the velocity broadening. In contrast, in the deep
dipping spectra, the values of \nhxabs\ tend to be poorly constrained
whereas \sigmav\ is well constrained. This indicates that the lines
are probably saturated and their \ews\ do not strongly depend on the
column density, but rather increase with velocity broadening. \exo\ is
very likely in a saturated regime already in the persistent
emission. This is indicated by the well constrained \sigmav\ value
together with the presence of absorption edges stronger than the
narrow absorption features.

We tried to model the spectral changes during dipping emission of the
studied LMXBs by changes only in the highly-ionized absorbers present
in these systems, but we could not obtain successful fits when the
deep dipping spectra are included, unless the \nh\ of the neutral
absorber is allowed to vary.  As an example, we could adequately model
the \nineteen\ persistent and dipping spectra when the \nh\ was fixed
for the persistent, Dip 1 and Dip 2 spectra (\rchisq\ = 1.36 for 1266
d.o.f.), but the fit quality became unacceptable when we fixed the
\nh\ of the deeper (Dip 3 to Dip 5) spectra, with structured residuals
evident $\approxlt$2~keV and a \rchisq\ = 2.44 for 1269 d.o.f. This
means that the spectral evolution requires changes in {\em both} the
ionized and the neutral absorbers.

For all the sources, the strongest absorption lines in the persistent
spectra are the \fetfive\ and \fetsix\ 1s-2p transitions except for
\twelve, where the absorber is the most highly ionized with \logxi\ =
4.2 $\pm$ 0.2 and only the Ly$\alpha$ and $\beta$ \fetsix\ absorption
features are evident. \exo\ and \fouru\ do not exhibit strong
absorption lines. Our analysis indicates that the \exo\ absorber is
significantly less ionized than in the other LMXBs studied with
\logxi\ = $2.45 \pm 0.02$.  This could indicate that the intervals
identified as ``persistent'' in Fig.~\ref{fig:lightcurves} are not
actually dip-free, and the source may be continuously dipping.  This
idea is supported by the large fraction of time that the source
appears to be dipping, the well constrained value of $\sigma _v$
during the ``persistent'' interval (Table~\ref{tab:bestfit-exo}) which
may indicate that the source is already in the saturated line regime
reached by the other sources only during dipping (see above), and by
the low luminosity when compared to the other sources with similar
values of $P{\rm _{orb}}$ (Table~\ref{tab:lmxb-prop}).  In contrast,
for \fouru\ the non-detection of an \fetfive\ absorption feature and
the evidence of one from \fetsix\ indicate a highly-ionized absorber,
similar to the one in \twelve. If the absorbing material is almost
completely ionized, no strong absorption lines will be
observed. Alternative explanations for the lack of deep absorption
lines in \fouru\ include an unusually low metallicity
\citep[e.g.,][]{1746:parmar99aa}, or a viewing angle outside of the
ionized plasma.

In order to investigate the effect of uncertainties in the ionizing
continuum on the spectral parameters for the ionized absorbers, we
performed fits to the \twelve\ spectra, using ionizing continua
generated with, and without, the 5.9~keV cutoff. Variations in $E {\rm
_c}$ are likely to dominate the uncertainties in the ionizing
continuum since the cutoff measurements cannot be obtained from fits
to the pn spectra, due to the restricted energy range of this
instrument (0.6--10~keV), whilst the power-law indices are directly
fit up to 10~keV. \twelve\ was chosen for this analysis since it has
the lowest cutoff energy in Table~\ref{tab:ioniz-cont}, except for
\fouru, which only exhibits shallow dips. The best-fit values of
\nhxabs\ and $\sigma _v$ are consistent within the errors for both
ionization continua, while \logxi\ decreases by $\sim$0.3 when the
cutoff is not included. Since this represents an extreme range of the
likely ionization continua, the actual uncertainties due to variations
in the cutoff of the ionizing continuum are likely to be significantly
smaller than this.

\begin{figure}[t!]
\centerline{\includegraphics[angle=0,width=0.60\textwidth]{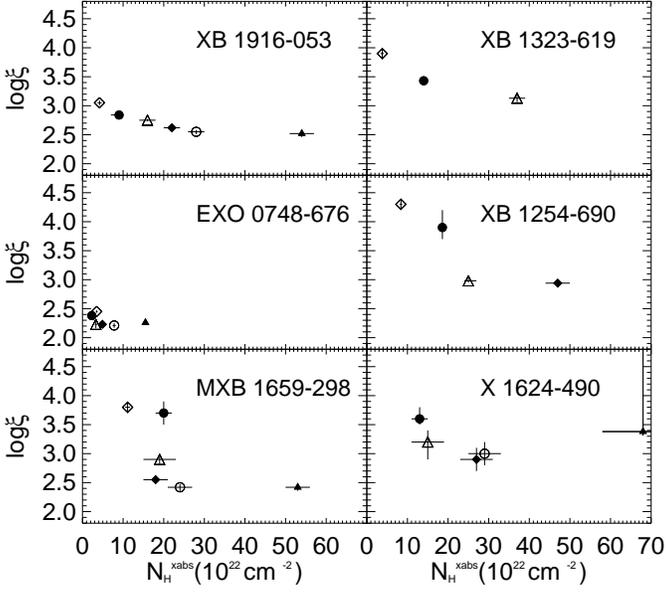}}
\vspace{0.2cm}
\caption{Evolution of \logxi\ and \nhxabs\ for the LMXBs studied in
this paper. Empty diamonds, filled circles, empty triangles, filled
diamonds, empty circles and filled triangles indicate the persistent
and Dip 1 to Dip 5 intervals, respectively.}
\label{fig:xil-nhx-ind}
\end{figure}

\begin{figure}[t!]
\centerline{\includegraphics[angle=0,width=0.60\textwidth]{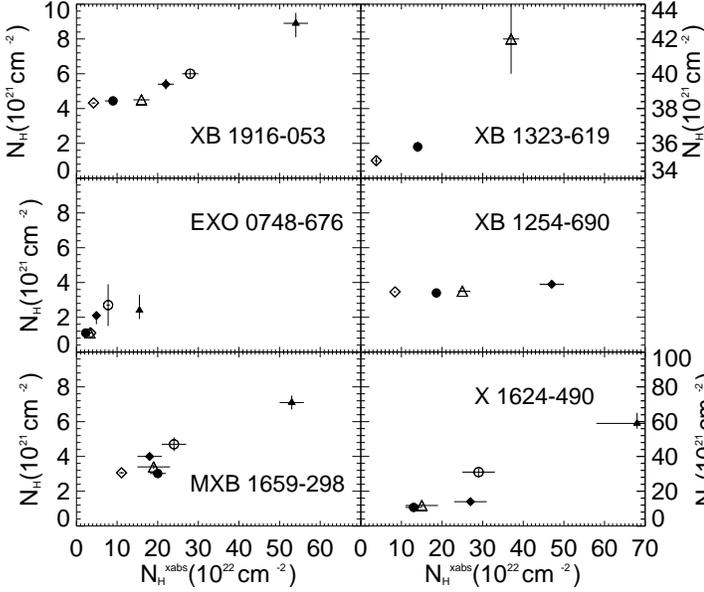}}
\vspace{0.2cm}
\caption{Evolution of the column densities of the neutral and ionized
absorbers, \nh, and \nhxabs, for the LMXBs studied in this
paper. Empty diamonds, filled circles, empty triangles, filled
diamonds, empty circles and filled triangles indicate the persistent
and Dip 1 to Dip 5 intervals, respectively.} \label{fig:nhx-nh-ind}
\end{figure}

The changes in absorber properties for each source as it evolves from
persistent to deep dipping do not strongly depend on the ionizing
continuum used.  Figure~\ref{fig:xil-nhx-ind} shows the evolution of
\xil\ and \nhxabs\ from persistent to the deepest dip intervals for
each source. A similar evolution of decreasing \xil\ and increasing
\nhxabs\ is observed for all sources, being the evolution stronger
from persistent to Dip 2 stages than between dips. For \exo, the
evolution of the highly-ionized absorber is unusually small compared
to the other sources. This could be explained if \exo\ were in a
continuous dipping state (see above).  Figure~\ref{fig:nhx-nh-ind}
shows the evolution of \nh, with respect to the ionized absorber,
\nhxabs, from persistent to deepest dip intervals for each source. The
increase of \nh\ is small in comparison to \nhxabs.  \twelve\ shows a
particularly small increase in \nh\ compared to the other
sources. This may indicate that we are viewing \twelve\ at an
inclination angle such that the line of sight is not obscured by
additional neutral material at any orbital phase and the dips result
{\em only} from additional obscuration by the ionized absorber.  Thus,
the source may be being viewed relatively far from the plane of the
accretion disk and only the ionized absorber significantly intercepts
the line of sight.  This special geometry and small changes in the
size of the ionized absorber may explain the remarkable complete
occasional disappearance of dipping activity from this source
\citep{1254:smale02apj, 1254:boirin03aa}. If this picture is correct,
the large changes in \nh\ observed from \bigdip\ during dips would
indicate that we are viewing this source very close to the plane of
the accretion disk.

\begin{figure}[t!]
\hspace{0.2cm}
\centerline{\hfill\includegraphics[angle=0,width=0.60\textwidth]{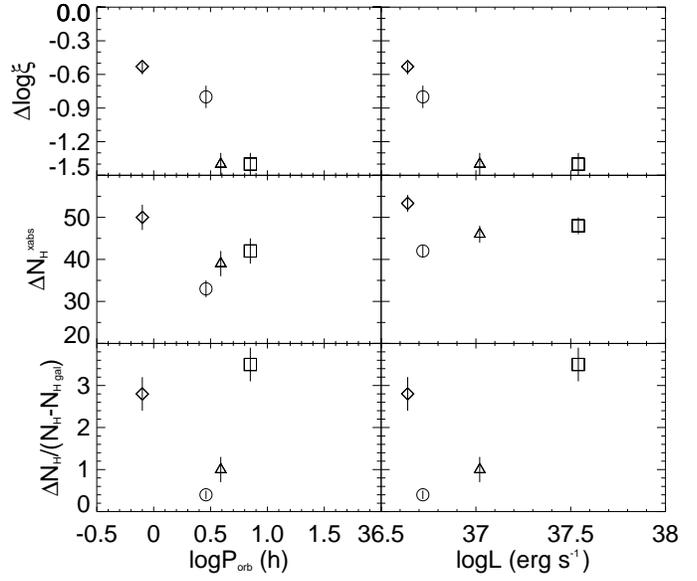}}
\vspace{0.2cm}
\caption{Changes of \logxi, \nhxabs\ and \nh\ from persistent to
deepest dip emission versus orbital period and luminosity for the
LMXBs studied here, except \exo\ and \bigdip\ (see text). The change
of \nh\ is expressed relative to the local persistent value of
\nh. Diamonds, circles, triangles and squares represent \nineteen,
\thirteen, \twelve\ and \mxb, respectively.} \label{fig:rchanges}
\end{figure}

Table~\ref{tab:delta} shows the changes in the properties of the
neutral and highly-ionized absorbers from persistent (Dip 1 for
\bigdip) to deepest dip phases.  We include in this table results for
\thirteen\ derived from the spectral fits presented in
\citet{1323:boirin05aa}.  These authors fixed the normalizations of
the dipping spectra to that of the persistent spectrum, thus a larger
change in the neutral absorber column may be necessary to account for
the changes when compared to fitting simultaneously the persistent and
dipping spectra, as was done for the other sources.  It is interesting
that the eclipsing binaries \exo\ and \mxb\ (together with the
non-eclipsing system \bigdip) show the largest change in \nh. This
suggests that the size of the change in \nh\ may be related to the
inclination angle.  This supports the idea that we are seeing \bigdip\
and \twelve\ very close to, and relatively far from, the planes of the
accretion disks. Alternative explanations for the large increase in
\nh\ for \bigdip\ include the high luminosity of this source,
uncertainties in modeling the dust scattering halo, or an absorber
rich in blobs of neutral material. Furthermore, \twelve\ and
\thirteen, which show the smallest relative changes of \nh, have also
the less deep dips, consistent with viewing the sources relatively far
from the disk plane.  We do not find a correlation between the dip
depth and the change in \nhxabs\ or \logxi. \exo\ and \mxb, which show
the deepest dips, have also the lowest values of \xil\ during the
deepest dips.  The change in \nhxabs\ is approximately the same for
all the sources except \exo. This is shown in Fig.~\ref{fig:rchanges}
by plotting the changes of the neutral and ionized absorbers versus
$P_{\rm orb}$ and $L$ of the studied sources. Note that we have
excluded \bigdip\ from this figure, whose persistent emission has not
been analyzed, and \exo, which may be continuously dipping.  In
contrast to \nhxabs, the change of \logxi\ appears to increase with
the source $P_{\rm orb}$ and $L$.

\begin{table*}
\caption{The persistent (Dip 1 for \bigdip) values of \nh\ (col. 2),
\nhxabs\ (col. 6) and \logxi\ (col. 8) and the changes in \nh\
(col. 3) and \nhxabs\ (col. 7) from persistent to the deepest dip
intervals observed for each source.  \nh$\rm _{gal}$ is the averaged
interstellar value for the 0\fdg5 region in the sky containing the
source \citep{dickey90araa}. $\Delta$\nh/(\nh$\rm _{pers}$-\nh$\rm
_{gal}$) is the relative change in \nh\ local to the source from
persistent to the deepest dip interval. Col. 9 shows the value of
\logxi\ during the deepest dip for each source.  \nh\ for \exo\ is
constrained to be $\ge 1.1 \times 10^{21}$~atom~cm$^{-2}$. All values
of \nh, \nhxabs\ and their changes are expressed in units of
$10^{22}$~atom~cm$^{-2}$.  }\begin{center}
\begin{tabular}{l@{\extracolsep{0.18cm}}c@{\extracolsep{0.17cm}}c@{\extracolsep{0.15cm}}c@{\extracolsep{0.15cm}}c@{\extracolsep{0.15cm}}c@{\extracolsep{0.15cm}}c@{\extracolsep{0.15cm}}c@{\extracolsep{0.18cm}}c@{\extracolsep{0.2cm}}c}
\hline \hline\noalign{\smallskip}
LMXB & \nh$\rm _{pers}$ & $\Delta$\nh\ & \nh$\rm_{gal}$ & $\Delta$\nh/ & \nhxabs$\rm _{pers}$ & $\Delta$\nhxabs\ & \logxi$\rm _{pers}$ & \logxi$\rm _{dip}$ & Dip \\
& & & & (\nh$\rm _{pers}$-\nh$\rm _{gal}$) & & & & & depth\\
\noalign{\smallskip\hrule\smallskip}
\nineteen\ & 0.432 $\pm$ 0.002  & 0.46 $\pm$ 0.07& 0.27 & 2.8 $\pm$ 0.4 & 4.2 $\pm$ 0.5 & 50 $\pm$ 3 & 3.05 $\pm$ 0.04 & 2.52 $^{+0.02}_{-0.06}$ & 80\% \\
\thirteen $^a$ & 3.50 $\pm$ 0.02 &  0.7 $\pm$ 0.2 & 1.57 & 0.4 $\pm$ 0.1 & 3.8 $\pm$ 0.4 & 33 $\pm$ 2 & 3.9 $\pm$ 0.1 & 3.13 $\pm$ 0.07 & 75\% \\
\exo\ & 0.11 &  0.13 $\,^{+0.09}_{-0.05}$ & 0.11 & $\infty$ & 3.5 $\pm$ 0.2 & 12.0 $\pm$ 0.5 & 2.45 $\pm$ 0.02 & 2.26 $\pm$ 0.03 & \llap{$>$}85\% \\
\twelve\ & 0.346 $\pm$ 0.002 &  0.04 $\pm$ 0.01 & 0.31 & 1.0 $\pm$ 0.3 & 8.4 $\pm$ 0.3 & 39 $\pm$ 3 & 4.3 $\pm$ 0.1 & 2.94 $\pm$ 0.05 & 50\% \\
\mxb\ & 0.306 $\pm$ 0.003 &  0.40 $\pm$ 0.04 & 0.19 & 3.5 $\pm$ 0.4 & 11.1 $\pm$ 0.6 & 42 $\pm$ 3 & 3.8 $\pm$ 0.1 & 2.42 $\,^{+0.02}_{-0.06}$ & \llap{$>$}85\% \\
\bigdip $^b$ & 10.7 $\pm$ 0.5  & 48$\,^{+6}_{-3}$ & 2.22 & 5.7 $^{+0.7}_{-0.4}$ & 13 $\pm$ 2  & 55 $\pm$ 9 & 3.6 $\pm$ 0.2 & $\ge $3.3 & 80\% \\
\noalign{\smallskip\hrule\smallskip}
\label{tab:hia}
\end{tabular}
\end{center}
\addtocounter{footnote}{-1}
\footnotetext{}{$^a$Values for \thirteen\ are derived from the spectral fits in \citet{1323:boirin05aa}.}\stepcounter{footnote}
\\
\footnotetext{}{$^b$The changes for \bigdip\ are calculated
between the Dip 1 and Dip 5 stages.} \label{tab:delta}
\end{table*}

\begin{figure}[t!]
\hspace{0.2cm}
\centerline{\hfill\includegraphics[angle=0,width=0.60\textwidth]{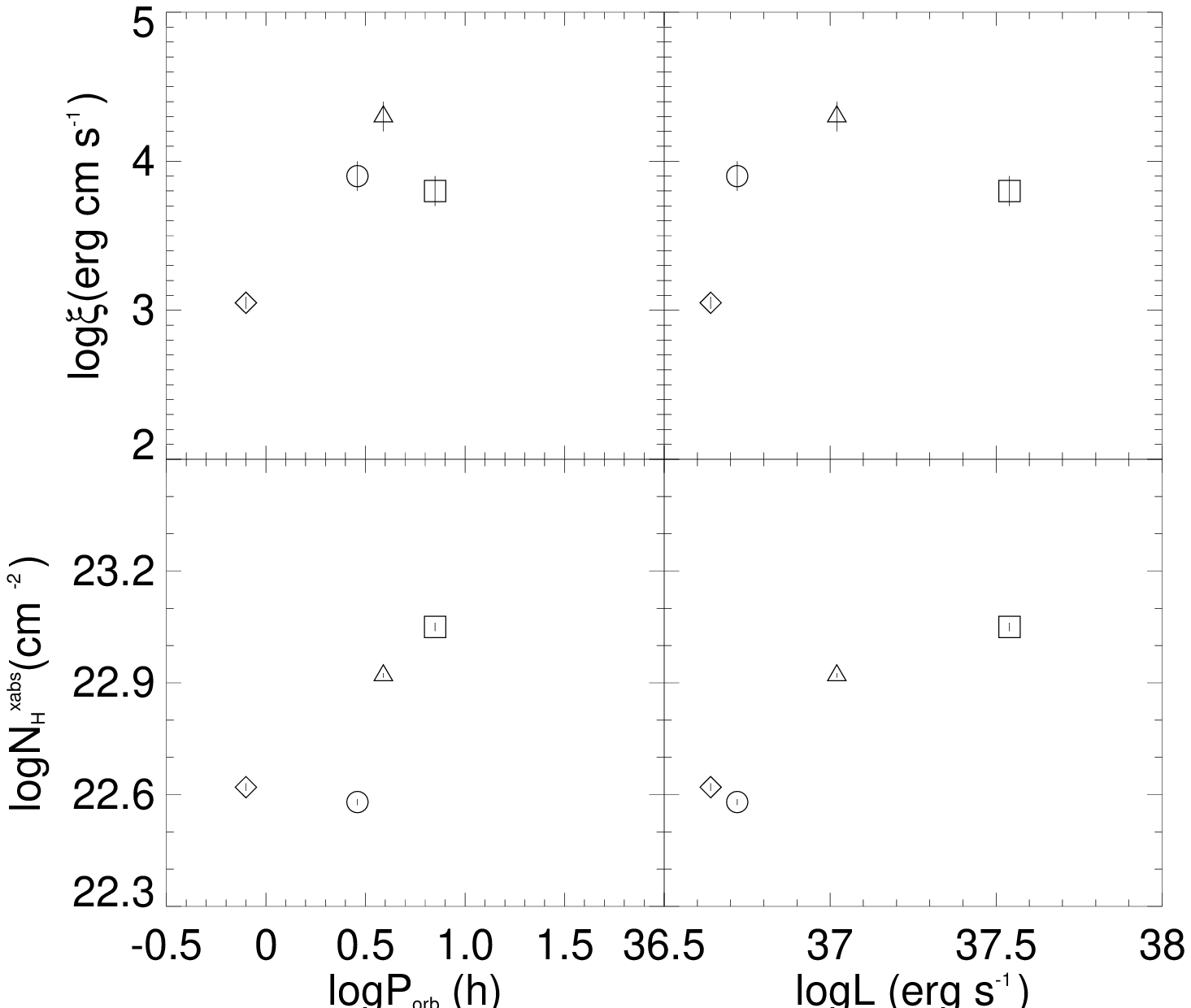}}
\vspace{0.2cm}
\caption{Values of \logxi\ and log\nhxabs\ for the ionized absorbers
during persistent emission versus $P_{\rm orb}$, and $L$, for the
LMXBs studied here, except \exo\ and \bigdip\ (see text). Diamonds,
circles, triangles and squares represent \nineteen, \thirteen,
\twelve\ and \mxb, respectively.} \label{fig:xil-period}
\end{figure}

It is interesting to compare the values of the ionization parameter
for the different sources. Since $\xi = L /n_{\rm e} \, r^{2}$, the
relation between $\xi$ and $L$ can provide information about the
distance between the absorbing material and the ionizing source, $r$.
Figure~\ref{fig:xil-period} shows the ionized absorber parameters
\logxi\ and \nhxabs\ as a function of $P_{\rm orb}$ and source
luminosity $L$ for the persistent emission. Note that we excluded
\bigdip\ and \exo\ from this figure (see above).  The \logxi\ and
\nhxabs\ of the absorbers in our sample appear to increase with
luminosity. Since the luminosity is proportional to the mass accretion
rate, this suggests that the column of the ionized absorber increases
with the mass accretion rate. \mxb\ has a lower \xil\ than the less
luminous \twelve. This indicates that either $n_{\rm e}$ or $r$ are
much larger for \mxb\ than for \twelve. Assuming a constant $n_{\rm
e}$, the evolution of \logxi\ with log$L$ suggests that for low $L$
sources, the change of $r$ is small compared to the $L$, while for
sources with $L$ $\approxgt$10$^{37}$~erg~s$^{-1}$, $r$ increases
faster than $L$.  There may also be a correlation between \logxi\ or
\nhxabs\ and $P_{\rm orb}$.  The increase of \logxi\ with log$(P_{\rm
orb})$ indicates that the system period is less important than the
luminosity of the system for the degree of ionization of the plasma,
since otherwise systems with small $P_{\rm orb}$, and correspondingly
smaller distances between the source and the absorber, should ionize
more effectively the absorbing material.

We can compare the thickness of the slab of ionized absorbing
material, $d$, with the distance between the ionizing source and the
slab, $r$.  Since $\xi = L /n_{\rm e} \, r^{2}$ and $n_{\rm
e}\sim$~\nhxabs/$d$, we can calculate the ratio, $d/r$, as
$\xi$~\nhxabs~$r/ \, L$.  We estimated a range of possible ratios for
persistent and deepest dip emission taking into account the
uncertainties in the determination of $r$ and $L$. We considered the
circularization radius as the minimum possible value for $r$ and
0.8$r_L$, where $r_L$ is the radius of the Roche lobe, as the maximum
possible value for $r$. We assumed that the accretion occurs via Roche
lobe overflow and calculated the radii from $P_{\rm orb}$ (eqs.~4.16
from \citealt{frank92apia}, and 4.18 from \citealt{shore94saas}). In
the determination of $L$ the largest uncertainty is given by the
uncertainty in the distance. We calculated an average value for the
distance from all the values given in the literature and considered an
uncertainty of $\pm$3 kpc for each source (see
Table~\ref{tab:slab_thick}).

We note that during dipping emission \exo\ and \mxb, which have
relatively cool absorbers (\logxi\ $\approxlt$ 2.5) and show deep and
sharp dips, have $d \, < \, r$, while the sources with more ionized
absorbers have $d \, \sim \, r$.  During persistent emission, all the
values of $d$ are consistent with $r$ (taking uncertainties into
account) except for \nineteen\ and \exo, which show small values of
$d/r$ ($\sim$ 0.04-0.6) and the smallest values of $\xi$. This may
indicate that the material responsible for the ionized absorption seen
during persistent emission is clumpy and located at the outer edge of
the accretion disk for \nineteen\ and \exo\ and distributed on a
significant fraction of the accretion disk for the other sources.

\begin{table}
\caption{Slab thickness of ionized absorbing material, $d$, with
respect to the size of the accretion disk, $r$, during persistent and
deepest dip emission for all the studied sources. Col.~2 shows
estimates of $r$ obtained from $P_{\rm orb}$ and col.~3 estimates of
the source distances (see text).  }\begin{center}
\begin{tabular}{lcccc}
\hline \hline\noalign{\smallskip}
LMXB & r & dist & (d/r)$\rm _{pers}$ & (d/r)$\rm _{dip}$ \\
& (10$^{10}$ cm) & (kpc) & & \\
\noalign{\smallskip\hrule\smallskip}
\nineteen\ & 1.1-2.0 & 6-12 & 0.07-0.5 & 0.3-2 \\
\thirteen\ & 1.5-4.2 & 7-13 & 0.5-5 & 0.9-8 \\
\exo\ & 1.7-4.9 & 5-11 & 0.04-0.6 & 0.1-2 \\
\twelve\ & 1.7-4.9 & 8-14 & 1-12 & 0.3-3 \\
\mxb\ & 2.1-7.0 & 10-16 & 0.4-3 & 0.07-0.6 \\ 
\bigdip & 3.8-13.3 & 12-18 & 0.3-2 & 0.8-6 \\
\noalign{\smallskip\hrule\smallskip}
\label{tab:hia}
\end{tabular}
\end{center}

\label{tab:slab_thick}
\end{table}

\section{Discussion}
\label{sec:discussion}

We have demonstrated that the complex changes in the
0.6--10~keV continuum and absorption lines during dips from
most of the LMXBs studied may be self-consistently understood
as resulting from large increases in
\nhxabs and decreases in \xil\ of a
highly-ionized absorber, together with much smaller increases in the \nh\
of a neutral absorber. These changes are similar to those found by
\citet{1323:boirin05aa} for \thirteen. We do not need to invoke
unusual abundances or partial coverage of an extended emission region
to account for these changes.

Narrow X-ray absorption lines were first detected from the
superluminal jet sources \gro\
\citep{1655:ueda98apj,1655:yamaoka01pasj} and \grs\
\citep{1915:kotani00apj,1915:lee02apj}. \gro\ has been observed to
undergo deep absorption dips \citep{kuulkers98apj} consistent with
observing the source at an inclination angle of
$60\degmark$--$75\degmark$
\citep[e.g.,][]{1655:orosz97apj,1655:vanderhooft98aa}. Recently {\it
Chandra} HETGS observations of the black hole candidates
\threethreenine, \xte\ and \seventeen\
\citep{gx339:miller04apj,1743:miller04apj} have revealed the presence
of variable, blue-shifted, highly-ionized absorption features which
are interpreted as evidence for outflows.  While \fetfive\ and
\fetsix\ features are present in the \seventeen\ spectrum,
\threethreenine\ and \xte\ show \oeight\ and \nenine\ or \netwo\
features from less ionized material.  These features suggest that a
warm absorber analogous to those seen in many Seyfert galaxies is
present in systems such as \threethreenine\ and \xte\
\citep{gx339:miller04apj}. In contrast, \citet{1743:miller04apj}
propose for \seventeen\ a highly-ionized absorber, which may be a
precursor to the cooler outflows observed in \threethreenine\ and
\xte.  XMM-Newton observations of the LMXB GX\thinspace13+1 revealed
absorption features due to \catwenty, \fetfive\ and \fetsix\
\citep{gx13:sidoli02aa} and an ionized outflow with a velocity of
$\sim$400~km~s$^{-1}$ was reported by \citet{gx13:ueda04apj}.
Blueshifted outflows have not been detected from any of the dipping
LMXBs \citep{1658:sidoli01aa, 1624:parmar02aa, 1254:boirin03aa,
1916:boirin04aa, 1323:boirin05aa}. However, these results are all
obtained with the EPIC which has a factor $\sim$4 poorer energy
resolution than the HETGS at 6~keV limiting the sensitivity to shifts
$\approxgt$1000~km~s$^{-1}$. Since dipping sources are simply normal
LMXBs viewed from close to the orbital plane, this also implies that
ionized absorbers, frequent in micro-quasars and AGNs
\citep[e.g.,][]{reynolds97mnras,blustin05aa}, are also a common
feature of LMXBs. Outside of the dips, the properties of the
absorption lines do not appear to vary strongly with orbital
phase. This suggests that the ionized plasma in LMXBs has a
cylindrical geometry with a maximum column density close to the plane
of the accretion disk.

The spectral changes during dips from LMXBs are often modeled
using the ``progressive covering'', or ``complex continuum''
approach \citep[e.g.,][]{1624:church95aa, 1323:barnard01aa}.
There the X-ray emission is assumed to originate from a point-like
blackbody, or disk-blackbody component, together with an extended
power-law component. This approach models the spectral changes
during dipping intervals by the partial and progressive covering
of the extended component by an opaque absorber. We have
self-consistently demonstrated that changes in the properties of
an ionized absorber
provide an alternative explanation for the overall spectral
changes during dips from all the dipping LMXBs studied by XMM-Newton.
Further investigations and particularly the high spectral
resolution observations of LMXBs expected from Astro-E2
should allow many of the features predicted by the ionized
absorber model to be characterized and outflowing plasmas in LMXBs
to be studied.

\begin{acknowledgements}
  Based on observations obtained with XMM-Newton, an ESA science
  mission with instruments and contributions directly funded by ESA
  member states and the USA (NASA).  M. D{\'i}az Trigo acknowledges an
  ESA Fellowship. SRON is supported financially by NWO, the
  Netherlands Organization for Scientific Research. We thank the
  anonymous referee for helpful comments.
\end{acknowledgements}


\bibliographystyle{aa}
\bibliography{mybib,extrabiblio}

\end{document}